\documentclass[fleqn,twoside]{article}
\usepackage{espcrc2}
\usepackage{graphicx,here,rotate,epsf,epsfig}
\usepackage{axodraw}
 
\def\ga{\mathrel{\raise.3ex\hbox{$>$\kern-.75em\lower1ex\hbox{$\sim$}}}}
\def\la{\mathrel{\raise.3ex\hbox{$<$\kern-.75em\lower1ex\hbox{$\sim$}}}}
\def\gev{{\rm \, Ge\kern-0.125em V}}
\def\tev{{\rm \, Te\kern-0.125em V}}
\def\beq{\begin{equation}}
\def\eeq{\end{equation}}

\def\m12{m_{1\!/2}}

\def\ga{\mathrel{\raise.3ex\hbox{$>$\kern-.75em\lower1ex\hbox{$\sim$}}}}
\def\la{\mathrel{\raise.3ex\hbox{$<$\kern-.75em\lower1ex\hbox{$\sim$}}}}
\def\gyr{{\rm \, G\kern-0.125em yr}}
\def\gev{{\rm \, Ge\kern-0.125em V}}
\def\tev{{\rm \, Te\kern-0.125em V}}
\def\beq{\begin{equation}}
\def\eeq{\end{equation}}

\def\m12{m_{1\!/2}}

\def\gappeq{\mathrel{\rlap {\raise.5ex\hbox{$>$}}
{\lower.5ex\hbox{$\sim$}}}}

\def\lappeq{\mathrel{\rlap{\raise.5ex\hbox{$<$}}
{\lower.5ex\hbox{$\sim$}}}}

\def\Toprel#1\over#2{\mathrel{\mathop{#2}\limits^{#1}}}

\newcommand{\AmS}{{\protect\the\textfont2
  A\kern-.1667em\lower.5ex\hbox{M}\kern-.125emS}}

\title{Sneutrino Inflation}

\title{SNEUTRINO INFLATION}
 
\author{John Ellis\address{Theory Division, Physics Department, CERN, 
Geneva, Switzerland}}

\begin{document}

\begin{abstract}

The seesaw model of neutrinos might explain the size, age, flatness and
near-homogeneity of the Universe via sneutrino inflation, as well as
explaining the origin of matter via leptogenesis. The sneutrino inflation
hypothesis makes specific, testable predictions for cosmic microwave
background observables, which are compatible with the first release of
data from WMAP, and for flavour-violating charged-lepton decays. In
particular, $\mu \to e \gamma$ should occur with a branching ratio very
close to the present experimental upper limit, whilst $\tau \to \mu
\gamma$ and $e \gamma$ should occur further below the present limits.
\vspace{1pc}
\end{abstract}

\maketitle

\begin{center}
CERN-PH-TH/2004-057 $\; \;$ {\tt hep-ph/0403247}
\end{center}
 
\begin{center}
{\it Invited talk at the Fujihara Seminar on Neutrino Mass and  
Seesaw Mechanism, KEK, Feb. 23-25, 2004} 
\end{center}

\section{INTRODUCTION}

What is the connection between the data on the cosmic microwave background
(CMB) from WMAP~\cite{WMAP} and searches for $\mu \to e \gamma$ and other
charged-lepton flavour-violating (LFV) processes? One may be provided by
sneutrino inflation, an idea first proposed by Murayama-san, Suzuki-san,
Yanagida-san and Yokoyama-san~\cite{MSYY1,MSYY2}. This appealing idea
languished in obscurity for over a decade, for reasons that are unknown to
me. Personally, although I have always been attracted towards inflationary
cosmology, I have been reluctant to embrace it wholeheartedly, largely
because of the absence of a convincing candidate for the inflaton field.
Only recently did it dawn upon me that the supersymmetric spin-zero
partner of one of the heavy singlet neutrinos in the seesaw might be a
suitable candidate. My interest was then stimulated by the first release
of data from the WMAP satellite~\cite{WMAP}, which were consistent with
many key inflationary predictions, such as the Gaussian nature and almost
scale invariance of the primordial density perturbations. Moreover, the
WMAP data excluded many rival inflationary models, such as those with a
simple quartic potential, for example.

Last year, Raidal, Yanagida-san and I~\cite{ERY} revived the sneutrino
inflation idea, calculated several inflationary observables such as the
scalar spectral index (tilt) and the magnitude of the tensor perturbations
relative to the scalar modes, showed they were consistent with the WMAP
data, used the model to constrain the seesaw parameters, and finally
calculated LFV processes showing, for example, that $\mu \to e \gamma$
might occur not far below the present experimental limit. More recently, 
Chankowski, Pokorski, Raidal, Turzynski and I~\cite{CEPRT} have explored 
in more detail the sneutrino inflation predictions for LFV processes. This 
talk is based on these two papers.

First, however, I summarize the basic features of
inflation~\cite{Inflation}. Then I recall the 18 parameters of the minimal
three-generation seesaw model~\cite{CI}, and how they appear in low-energy
LFV observables as well as neutrino oscillations and
leptogenesis~\cite{EHLR,ERaidal}. Then I compare the predictions of 
sneutrino
inflation~\cite{ERY} with WMAP data~\cite{WMAP} and recall that the
gravitino problem~\cite{gravitino} favours non-thermal scenarios for
leptogenesis~\cite{ERY}. Finally, I dissect~\cite{CEPRT} the predictions
of sneutrino inflation for various LFV processes~\cite{ERY}, showing how
they follow from the (near) decoupling of the sneutrino inflaton, which is
motivated by the gravitino problem and some approaches to neutrino masses 
within the seesaw model~\cite{Chank}.

\section{SUMMARY OF INFLATIONARY COSMOLOGY}

The basic idea of cosmological inflation~\cite{Guth} is that, at some 
early epoch in the history of
the Universe, its energy density may have been dominated by an almost
constant term:

\begin{equation}
\left( { {\dot a} \over a} \right)^2 \; = \; {8 \pi G_N \rho \over 3} - {k
\over a^2}: \; \; \rho \; = \; V,
\label{infV}
\end{equation}
leading to a phase of near-exponential de Sitter expansion.
It is easy to see that the second (curvature) term in (\ref{infV}) rapidly
becomes negligible, and that
\begin{equation}
a \; \simeq \; a_I e^{Ht}: \; \; H \; = \; \sqrt{ {8 \pi G_N \over 3} V }
\label{expV}
\end{equation}
during this inflationary expansion, if $V$ is really constant.

In this case, the {\it horizon} would also have expanded (near-)
exponentially, so that the entire visible Universe might have been within
our pre-inflationary horizon:
\begin{equation}
a_H \; \simeq \; a_I e^{H \tau} \; \gg \; c \tau,
\label{newhorizon}
\end{equation}
where $H \tau$ is the number of e-foldings during inflation.
This would have enabled our observable universe to appear (almost) 
homogeneous. Since the $- {k \over a^2}$ term in (\ref{infV})
becomes negligible, the Universe may now appear almost {\it flat} with
$\Omega_{tot} \simeq 1$. However, as we see later, perturbations during
inflation generate a small deviation from unity: $|\Omega_{tot} - 1 |
\simeq 10^{-5}$. Following inflation, the conversion of the inflationary
vacuum energy into
particles reheats the Universe, filling it with the required entropy. 
Finally, the closest pre-inflationary monopole or gravitino is
pushed away, further than the origin of the CMB, by the exponential
expansion of the Universe.

The above description is quite classical. In fact, one should expect
quantum fluctuations in the initial value of the inflaton field $\phi$,   
which would cause the inflationary expansion to be slightly
inhomogeneous~\cite{Inflation}, with different parts of the Universe 
expanding at
slightly different rates. These quantum
fluctuations would give rise to a Gaussian random field of 
density perturbations
with similar magnitudes on different scale sizes, just as the
astrophysicists have long wanted~\cite{HZ}. The magnitudes of these 
perturbations
would be linked to the value of the effective potential during inflation,
and would be visible in the CMB as adiabatic temperature 
fluctuations~\cite{Inflation}:
\begin{equation}
{\delta T \over T } \; \sim \; {\delta \rho \over \rho} \; \sim \; \mu^2
G_N,
\label{deltaT}
\end{equation}
where $\mu \equiv V^{1/4}$ is a typical vacuum energy scale during
inflation. Consistency with the cosmic microwave background (CMB)
data from COBE {\it et al}., that find $\delta T / T \simeq 10^{-5}$, is
obtained if
\begin{equation}
\mu \; \simeq \; 10^{16}~{\rm GeV},
\label{iscale}
\end{equation}
comparable with the GUT scale.

One example of such a scenario is chaotic inflation~\cite{chaos},
according to which there is no special structure in the effective
potential $V(\phi)$, which might be a simple power $V \sim \phi^n$ or
exponential $V \sim e^{\alpha \phi}$. In this scenario, any given region
of the Universe is assumed to start with some random value of the inflaton
field $\phi$ and hence the potential $V(\phi)$, which decreases
monotonically to zero. The equation of motion of the inflaton field is
\begin{equation}
{\ddot {\phi}} \; + \; 3 H {\dot \phi} \; + \; V^\prime(\phi) \; = \; 0,
\label{motion}
\end{equation}
and (our part of) the Universe undergoes
sufficient expansion if the initial value of $V(\phi)$ is large enough,
and the potential flat enough.
The first term in (\ref{motion}) is assumed to be negligible, in which
case the equation of motion is dominated by the second (Hubble drag) term,
and one has
\begin{equation}
{\dot \phi} \; \simeq \; - {V^\prime \over 3 H}.
\label{roll}  
\end{equation}
In this slow-roll approximation, $\phi$ rolls slowly down the potential if
\begin{eqnarray}
\epsilon \; &\equiv& \; {1 \over 2} M_P^2 \left( {V^\prime \over V}
\right)^2, \; 
\eta \; \equiv M_P^2 \left( {V^{\prime\prime} \over V} \right)~, \\ \nonumber
\xi \; &\equiv& \;  M_P^4 \left( {V V^{\prime\prime\prime} \over V^2} 
\right)
\label{epseta}
\end{eqnarray}
are all $\ll 1$,
where $M_P \equiv 1/\sqrt{8 \pi G_N} \simeq 2.4 \times 10^{18}$~GeV.
Various observable quantities can then be expressed in terms of
$\epsilon, \eta$
and $\xi$~\cite{Inflation}, including the spectral index for scalar 
density perturbations:
\begin{equation}
n_s \; = \; 1 \; - \; 6 \epsilon \; + \; 2 \eta,
\label{index}
\end{equation}  
the ratio of scalar and tensor perturbations at the quadrupole scale:
\begin{equation}
r \; \equiv {A_T \over A_S} = 16 \epsilon,
\label{scalartensor}
\end{equation}
the spectral index of the tensor perturbations:
\begin{equation}
n_T \; = \; -2 \epsilon,
\label{tensoridex}
\end{equation}
and the running parameter for the scalar spectral index:
\begin{equation}
{d n_s \over d {\rm ln} k} \; = \; {2 \over 3} \left[ \left( n_s - 1
\right)^2 -
4 \eta^2 \right] + 2 \xi.
\label{running}
\end{equation}  
The amount $e^N$ by which the Universe expanded during inflation is also
controlled~\cite{Inflation} by the slow-roll parameter $\epsilon$:
\begin{equation}
e^N : \; N \; = \; \int H dt \; = \; {2 \sqrt{\pi} \over m_P }
\int^{\phi_{final}}_{\phi_{initial}} {d \phi \over \sqrt{\epsilon (\phi)}
}.
\label{expansion}
\end{equation}
In order to explain the size of a feature in the observed Universe, one
needs:
\begin{eqnarray}
N \; & = &\; 62 - {\rm ln} {k \over a_0 H_0} - {\rm ln} {10^{16} {\rm GeV}
\over V_k^{1/4}} + {1 \over 4} {\rm ln} {V_k \over V_e} \nonumber \\
&-& {1 \over 3}
{\rm ln} {V_e^{1/4} \over \rho_{RH}^{1/4}} 
 \sim  50 \; {\rm to} \; 70,
\label{amount}
\end{eqnarray} 
where $k$ characterizes the size of the feature, $V_k$ is the magnitude of
the inflaton potential when the feature left the horizon, $V_e$ is the  
magnitude of the inflaton potential at the end of inflation, and
$\rho_{RH}$ is the density of the Universe immediately following
reheating after inflation.

As an example of the above general slow-roll theory, let us consider
chaotic inflation~\cite{chaos} with a $V = {1 \over 2} m^2
\phi^2$ potential, as appears in the sneutrino inflation
model~\cite{ERY} discussed later. In this model,
the conventional slow-roll inflationary parameters are
\begin{equation}
\epsilon \; = \; {2 M_P^2 \over \phi_I^2}, \;
\eta \; = \; {2 M_P^2 \over \phi_I^2}, \;
\xi \; = \; 0,
\label{slowroll2}
\end{equation}
where $\phi_I$ denotes the {\it a priori} unknown inflaton field value  
during inflation at a typical CMB scale $k$. The overall scale of the
inflationary potential is normalized by the WMAP data on density
fluctuations:
\begin{eqnarray}
\Delta_R^2 &=& {V \over 24 \pi^2 M_P^2 \epsilon} = 2.95 \times 10^{-9} A~,\\ \nonumber
A &=& 0.77 \pm 0.07,
\label{normn}
\end{eqnarray}
yielding
\begin{eqnarray}
V^{1 \over 4} &= &M_P ^4\sqrt{\epsilon \times 24 \pi^2 \times 2.27 \times
10^{-9}} \\ \nonumber
& =& 0.027 M_P \times \epsilon^{1 \over 4},
\label{WMAP}  
\end{eqnarray}   
corresponding to
\begin{equation}
m^{1 \over 2} \phi_I \; = \; 0.038 \times M_P^{3 \over 2}
\label{onecombn}
\end{equation}
in any simple chaotic $\phi^2$ inflationary model.
The above expression (\ref{amount}) for the number of
e-foldings after the
generation of the CMB density fluctuations observed by COBE
could be as low as $N \simeq 50$ for a reheating temperature $T_{RH}$ as
low as $10^6$~GeV. In the $\phi^2$ inflationary model, this value of $N$
would imply
\begin{equation}
N \; = {1 \over 4} {\phi^2_I \over M_P^2} \simeq \; 50,
\label{fixN}
\end{equation}
corresponding to 
\begin{equation}
\phi^2_I \simeq 200 \times M_P^2.
\label{fixphi}
\end{equation}  
Inserting this requirement into the WMAP normalization condition
(\ref{WMAP}), we find~\cite{ERY} the following required mass for any 
inflaton with a simple quadratic potential:
\begin{equation}
m \; \simeq 2 \times 10^{13}~{\rm GeV}.
\label{phimass}
\end{equation}
This is comfortably within the range of heavy
singlet (s)neutrino masses usually considered, namely $m_N \sim 10^{10}$
to $10^{15}$~GeV, motivating the sneutrino inflation model
discussed below.

Is this simple $\phi^2$ model compatible with the WMAP data?
It predicts the following values for the
primary CMB observables~\cite{ERY}: the scalar spectral index
\begin{equation}
n_s \; = \; 1 - {8 M_P^2 \over \phi^2_I} \simeq 0.96,
\label{ns}
\end{equation}
the tensor-to scalar ratio
\begin{equation}
r \; = \; {32 M_P^2 \over \phi^2_I} \; \simeq \; 0.16,
\label{r}
\end{equation}
and the running parameter for the scalar spectral index:
\begin{equation}
{d n_s \over d {\rm ln} k} \; = \;
{32 M_P^4 \over \phi^4_I}  \simeq  8 \times
10^{-4}.
\label{values}
\end{equation}
The value of $n_s$ extracted from WMAP data depends whether, for example,
one combines them with other CMB and/or large-scale structure data.
However, the $\phi^2$ model value $n_s \simeq 0.96$ appears to
be compatible with the data at the 1-$\sigma$ level~\cite{WMAP}. The 
$\phi^2$ model value $r\simeq 0.16$ for the relative tensor strength is
also compatible with the WMAP data. In fact, we note that the favoured
individual values for $n_s, r$ and ${d n_s / d {\rm ln} k}$ reported in an
independent analysis~\cite{realbarger} all coincide with the
$\phi^2$ model values, within the latter's errors!

One of the most interesting features of the WMAP analysis is the
possibility that ${d n_s / d {\rm ln} k}$ might differ from
zero~\cite{WMAP}. The $\phi^2$ model value ${d n_s / d {\rm ln} k} \simeq
8 \times 10^{-4}$ derived above is negligible compared with the WMAP
preferred value and its uncertainties. However, ${d n_s / d {\rm ln} k} =
0$ still appears to be compatible with the WMAP analysis at the 2-$\sigma$
level or better, so we do not regard this as a death-knell for the
$\phi^2$ model. On the other hand, all higher-order power-law potentials
$V \sim \phi^n: n > 2$ do seem to be excluded by the WMAP 
data~\cite{WMAP}.

\section{PARAMETERS IN THE SEESAW MODEL}

A generic seesaw model has a mass matrix~\cite{seesaw}:
\begin{eqnarray}
\left( \nu_L, N\right) \left(
\begin{array}{cc}
0 & M_{D}\\
M_{D}^{T} & M
\end{array}
\right)
\left(
\begin{array}{c}
\nu_L \\
N
\end{array}
\right),
\label{seesaw}
\end{eqnarray}
where each of the entries should be understood as a matrix in generation
space. In order to provide the two measured differences in neutrino
masses-squared, there must be at least two non-zero masses, and hence at
least two heavy singlet neutrinos $N_i$~\cite{Frampton,Morozumi}.
Presumably, all three light neutrino masses are non-zero, in which case
there must be at least three $N_i$. This is indeed what happens in simple
GUT models such as SO(10), but some models~\cite{fSU5} have more singlet
neutrinos~\cite{EGLLN}. Here, for simplicity we consider just
three $N_i$.

The effective mass matrix for light neutrinos in the seesaw model may be
written as:
\begin{equation}
{M}_\nu \; = \; Y_\nu^T {1 \over M} Y_\nu v^2,
\label{seesawmass}
\end{equation}
where we have used the relation $m_D = Y_\nu v$ with $v \equiv \langle 0
\vert H \vert 0 \rangle$. Taking $m_D \sim m_q$ or $m_\ell$ and requiring
light neutrino masses $\sim 10^{-1}$ to $10^{-3}$~eV, we find that heavy
singlet neutrinos weighing $\sim 10^{10}$ to $10^{15}$~GeV seem to be   
favoured, comparable with the $\phi^2$ inflaton mass inferred above CMB 
data.

It is convenient to work in the field basis where the charged-lepton   
masses $m_{\ell^\pm}$ and the heavy singlet-neutrino mases $M$ are real and
diagonal. The seesaw neutrino mass matrix ${M}_\nu$
(\ref{seesawmass}) may then be diagonalized by a unitary transformation 
$U$:
\begin{equation}
U^T {M}_\nu U \; = \; {M}_\nu^d. 
\label{diag}
\end{equation}
This diagonalization is reminiscent of that required for the quark mass
matrices in the Standard Model. In that case, it is well known that one 
can redefine the phases of the quark fields~\cite{EGN} so that the mixing
matrix $U_{CKM}$ has just one CP-violating phase~\cite{KM}. However, in the 
neutrino case,
there are fewer independent field phases, and one is left with 3
physical CP-violating parameters:
\begin{equation}
U \; = \; U_\nu P_0: \; P_0 \equiv {\rm Diag} \left(
e^{i\phi_1},
e^{i\phi_2}, 1 \right).
\label{MNSP}
\end{equation}  
Here $U_\nu$ is the light-neutrino mixing matrix
first considered by Maki, Nakagawa and Sakata (MNS)~\cite{MNS}, and $P_0$
contains 2 CP-violating phases $\phi_{1,2}$ that are in 
principle observable at low
energies, e.g., in neutrinoless double-$\beta$ decay. The MNS matrix 
describing neutrino oscillations may be written as
\begin{eqnarray}
U_\nu &= & \left(
\begin{array}{ccc}
c_{12} & s_{12} & 0 \\
- s_{12} & c_{12} & 0 \\
0 & 0 & 1   
\end{array}
\right)
\left(
\begin{array}{ccc}
1 & 0 & 0 \\
0 & c_{23} & s_{23} \\
0 & - s_{23} & c_{23}
\end{array}
\right) \times \nonumber \\
&\times& \left(
\begin{array}{ccc}
c_{13} & 0 & s_{13} \\
0 & 1 & 0 \\
- s_{13} e^{- i \delta} & 0 & c_{13} e^{- i \delta}
\end{array}
\right),
\label{MNSmatrix}
\end{eqnarray}
where $c_{ij} \equiv \cos \theta_{ij}, s_{ij} \equiv \sin \theta_{ij}$.

The CP-violating phase $\delta$ could in principle be measured by 
comparing the oscillation   
probabilities for neutrinos and antineutrinos and computing the
CP-violating asymmetry~\cite{DGH}:
\begin{eqnarray}
&&P \left( \nu_e \to \nu_\mu \right) - P \left( {\bar \nu}_e \to
{\bar \nu}_\mu \right) \;  = \; \nonumber \\
&&16 s_{12} c_{12} s_{13} c^2_{13} s_{23} c_{23} \sin \delta \nonumber \\
&& \sin \left( {\Delta m_{12}^2 \over 4 E} L \right)
\sin \left( {\Delta m_{13}^2 \over 4 E} L \right)
\sin \left( {\Delta m_{23}^2 \over 4 E} L \right).\nonumber \\ 
\label{CPV}
\end{eqnarray}
The measurement of $\delta$ is the Holy Grail of neutrino-oscillation 
physics, but does it have anything to do with leptogenesis, as has often 
been hoped? The answer is `no', unless one makes some supplementary 
hypothesis~\cite{Frampton,Morozumi}.

\begin{figure*}
\hspace{1cm}
\begin{picture}(400,300)(-200,-150)
\Oval(0,0)(30,60)(0)
\Text(-25,13)[lb]{ ${\bf Y_\nu}$  ,  ${\bf M_{N_i}}$}
\Text(-40,-2)[lb]{{\bf 15$+$3 physical}}
\Text(-30,-15)[lb]{{\bf parameters}}
\EBox(-70,90)(70,150)
\Text(-55,135)[lb]{{\bf Seesaw mechanism}}
\Text(-8,117)[lb]{${\bf {M}_\nu}$}
\Text(-65,100)[lb]{{\bf 9 effective parameters}}
\EBox(-200,-140)(-60,-80)
\Text(-167,-95)[lb]{{\bf Leptogenesis}}
\Text(-165,-113)[lb]{ ${\bf Y_\nu Y_\nu^\dagger}$ , ${\bf M_{N_i}}$}
\Text(-175,-130)[lb]{{\bf 9$+$3 parameters}}
\EBox(60,-140)(200,-80)
\Text(80,-95)[lb]{{\bf Renormalization}}
\Text(95,-113)[lb]{${\bf Y_\nu^\dagger L Y_\nu}$ , ${\bf M_{N_i}}$}
\Text(80,-130)[lb]{{\bf 13$+$3 parameters}}
\LongArrow(0,30)(0,87)
\LongArrow(-45,-20)(-130,-77)
\LongArrow(45,-20)(130,-77)
\end{picture}
\caption{
Roadmap for the physical observables derived from $Y_\nu$ and
$N_i$~\protect\cite{ERaidal}.}
\label{fig:map}
\end{figure*}
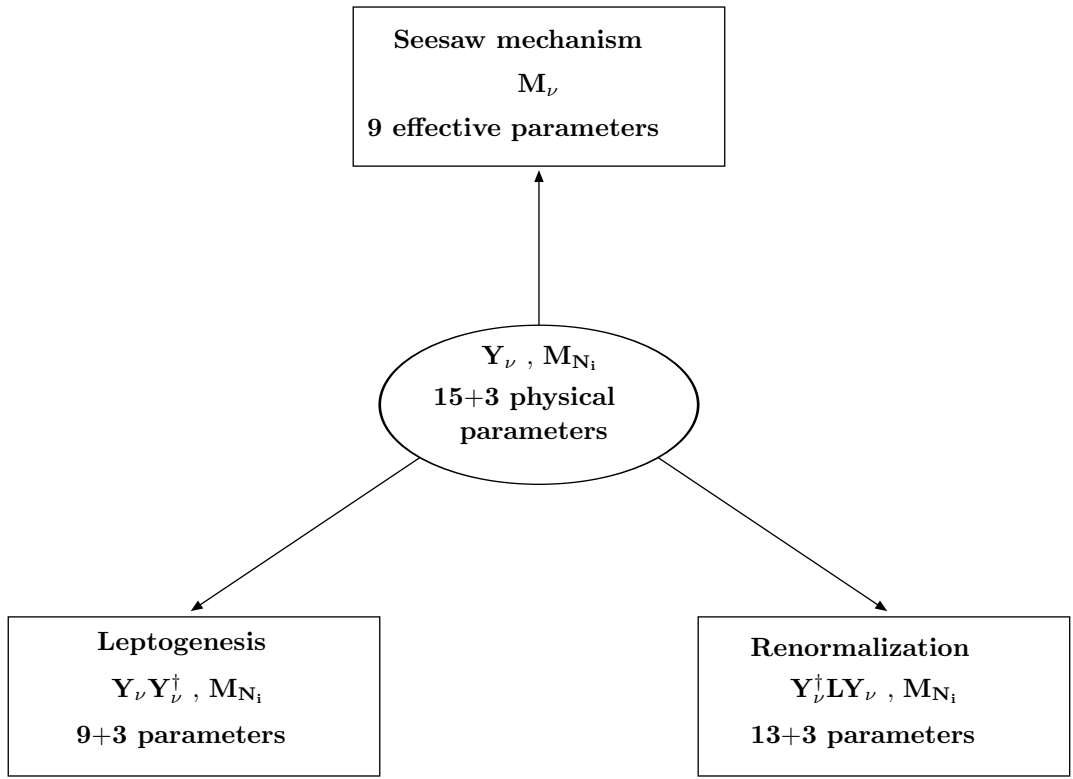

We have seen above that the effective low-energy mass matrix for the light
neutrinos contains 9 parameters, 3 mass eigenvalues, 3 real mixing angles
and 3 CP-violating phases. However, these are not all the parameters in
the minimal seesaw model. As shown in Fig.~\ref{fig:map}, this model has a
total of 18 parameters~\cite{CI,EHLR}. The additional 9 parameters
comprise the 3 masses of the heavy singlet `right-handed' neutrinos $M_i$,
3 more real mixing angles and 3 more CP-violating phases.

To see how the extra 9 parameters appear~\cite{EHLR}, we reconsider the
full lepton sector, assuming that we have diagonalized the charged-lepton 
mass
matrix:
\begin{equation}
\left( Y_\ell \right)_{ij} \; = \; Y^d_{\ell_i} \delta_{ij},
\end{equation}
as well as that of the heavy singlet neutrinos:
\begin{equation}
M_{ij} \; = M^d_i \delta_{ij}.
\label{diagM}
\end{equation}
We can then parametrize the neutrino Dirac coupling matrix $Y_\nu$ in
terms of its real and diagonal eigenvalues and unitary rotation matrices:
\begin{equation}
Y_\nu \; = \; Z^* Y^d_{\nu_k} X^\dagger,
\label{diagYnu}
\end{equation}
where $X$ has 3 mixing angles and one CP-violating phase, just like the
CKM matrix, and we can write $Z$ in the form
\begin{equation}
Z \; = \; P_1 {\bar Z} P_2,
\label{PZP}
\end{equation}
where ${\bar Z}$ also resembles the
CKM matrix, with  3 mixing angles and one CP-violating phase, and the
diagonal matrices $P_{1,2}$ each have two CP-violating phases:
\begin{equation}
P_{1,2} \; = \; {\rm Diag} \left( e^{i\theta_{1,3}}, e^{i\theta_{2,4}}, 1
\right).
\label{PP}
\end{equation}  
In this parametrization, we see explicitly that the neutrino
sector has 18 parameters: the 3 heavy-neutrino mass eigenvalues
$M^d_i$, the 3 real eigenvalues of $Y^D_{\nu_i}$, the $6 = 3 + 3$ real
mixing angles in $X$ and ${\bar Z}$, and the $6 = 1 + 5$ CP-violating  
phases in $X$ and ${\bar Z}$~\cite{EHLR}.   

As illustrated in Fig.~\ref{fig:map}, many of the extra seesaw parameters 
may be observable via
renormalization in supersymmetric models~\cite{BM,DI,EHLR,EHRS,EHRS2}, 
which
may generate observable rates for flavour-changing lepton decays such as
$\mu \to e \gamma, \tau \to \mu \gamma$ and $\tau \to e \gamma$, and
CP-violating observables such as electric dipole moments for the electron
and muon. In leading order, the extra seesaw parameters contribute to the
renormalization of soft supersymmetry-breaking masses, via a combination
which depends on just 1 CP-violating phase. However, two more phases
appear in higher orders, when one allows the heavy singlet neutrinos to be
non-degenerate~\cite{EHRS}. Some of these extra parameters may also have  
controlled the generation of matter in the Universe via
leptogenesis~\cite{FY}.

As discussed by many speakers here, the decays of the heavy singlet
neutrinos $N$ provide a mechanism for generating the baryon asymmetry of
the Universe, namely leptogenesis~\cite{FY}. In the presence of C and CP
violation, the branching ratios for $N \to {\rm Higgs} + \ell$ may differ
from that for $N \to {\rm Higgs} + {\bar \ell}$, producing a net lepton
asymmetry in the very early Universe. This is then transformed (partly)
into a quark asymmetry by non-perturbative electroweak sphaleron
interactions during the period before the electroweak phase transition.

The total decay rate of a heavy neutrino $N_i$ may be written in the   
form
\begin{equation}
\Gamma_i \; = \; {1 \over 8 \pi} \left( Y_\nu Y^\dagger_\nu \right)_{ii}
M_i.
\label{gammai}
\end{equation}
One-loop CP-violating diagrams involving the exchange of a heavy
neutrino $N_j$ would generate an asymmetry in $N_i$ decay of the form:
\begin{equation}
\epsilon_{ij} \; = \; {1 \over 8 \pi} {1 \over \left( Y_\nu Y^\dagger_\nu
\right)_{ii}} {\rm Im} \left( \left( Y_\nu Y^\dagger_\nu \right)_{ij}   
\right)^2 f \left( {M_j \over M_i} \right),
\label{epsilon}
\end{equation}
where $f ( M_j / M_i )$ is a known kinematic function.

\begin{figure*}
\centerline{\epsfxsize = 0.45\textwidth \epsffile{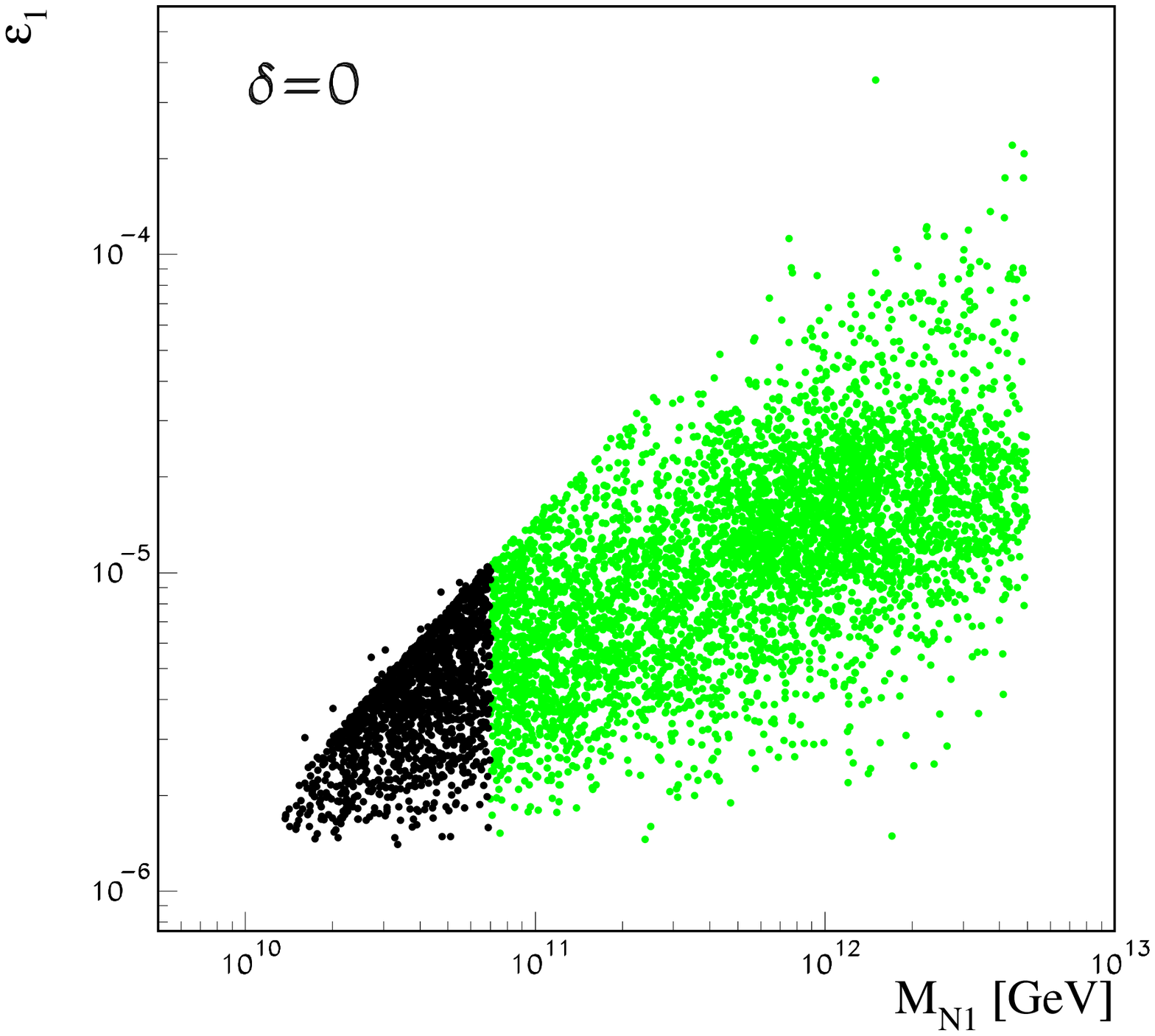}
\hfill \epsfxsize = 0.45\textwidth \epsffile{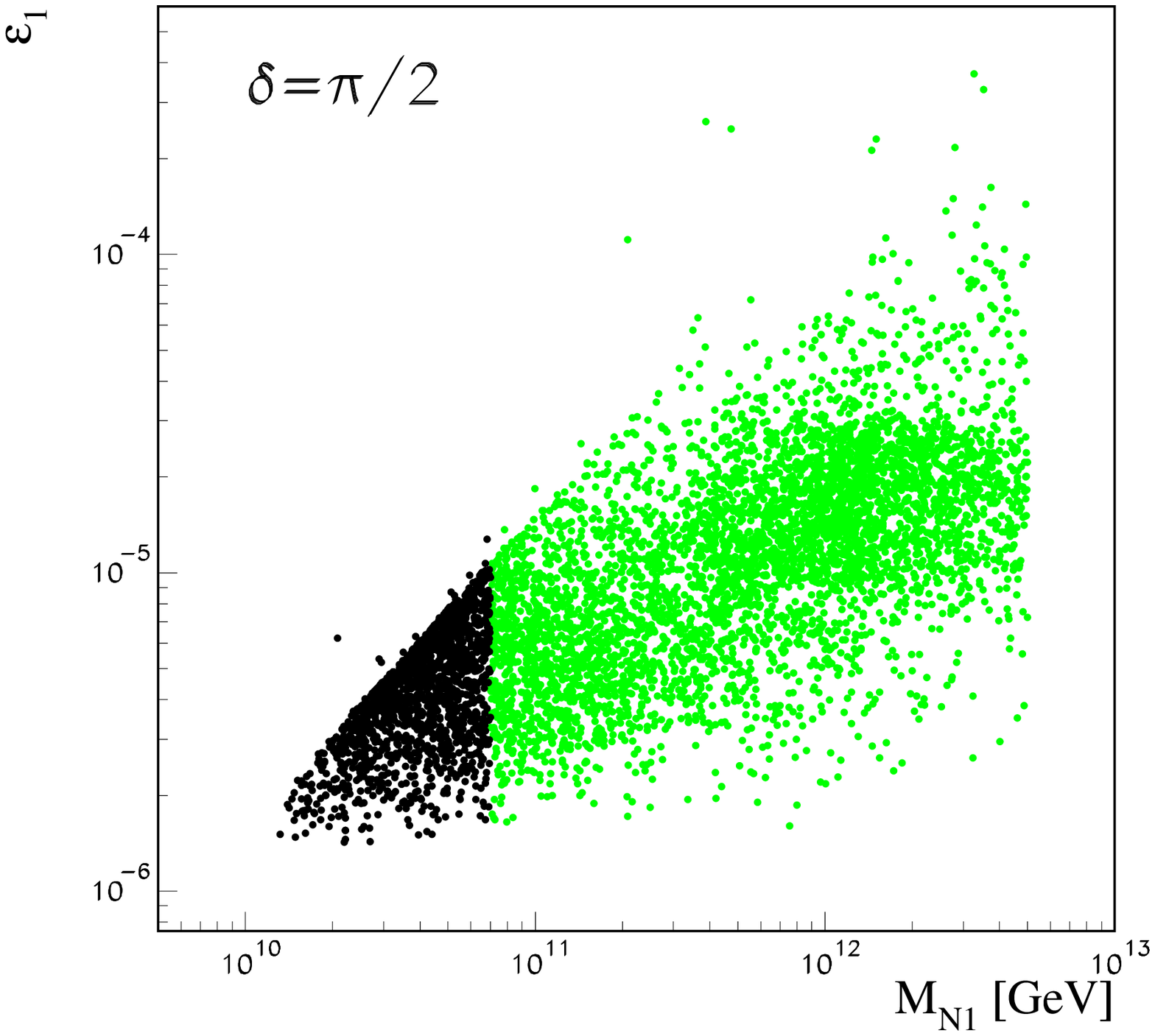}}
\caption{
Comparison of the CP-violating asymmetries in the decays of heavy singlet
neutrinos giving rise to the cosmological baryon asymmetry via
leptogenesis (left panel) without and (right panel) with maximal
CP violation in neutrino oscillations~\protect\cite{ERaidal}. They are
indistinguishable.}
\vspace*{0.5cm}
\label{fig:nodelta}
\end{figure*}

Thus we see that leptogenesis~\cite{FY} is proportional to the product 
$Y_\nu Y^\dagger_\nu$, which depends on 13 of the real parameters and 3
CP-violating phases.
However, as seen in Fig.~\ref{fig:nodelta}, the
amount of the leptogenesis asymmetry is explicitly independent of the
CP-violating phase $\delta$ that is measurable in neutrino
oscillations~\cite{ERaidal}. The basic reason for this is that one makes a
unitary sum over all the light lepton species in evaluating the
asymmetry $\epsilon_{ij}$. This does not mean that measuring $\delta$ is 
of no interest for leptogenesis: if it is found to be non-zero, CP violation
in the lepton sector - one of the key ingredients in leptogenesis - will
have been established. On the other hand, the phases responsible directly
for leptogenesis are not related directly to $\delta$, though they may 
contribute to other low-energy CP-violating observables such as the 
electric dipole moments of leptons.

Let us now discuss the renormalization of soft supersymmetry-breaking
parameters $m_0^2$ and $A$ in more detail, assuming that the input values
at the GUT scale are flavour-independent~\cite{BM}. If they are not, there 
will be
additional sources of flavour-changing processes, beyond those discussed
here~\cite{FCNI,Masiero}. In the leading-logarithmic
approximation, and assuming for simplicity degenerate heavy singlet 
neutrinos, one finds the following radiative corrections to the soft
supersymmetry-breaking terms for sleptons:

\begin{eqnarray}
\left( \delta m^2_{\tilde L} \right)_{ij} & \ni &
- { 1 \over 8 \pi^2} \left( 3 m_0^2 + A_0^2 \right) \left( Y_\nu^\dagger 
Y_\nu \right)_{ij}\nonumber \\
&& {\rm Ln} \left( {M_{GUT} \over M} \right), \nonumber \\
\left( \delta A_{\tilde L} \right)_{ij} & \ni &
- { 1 \over 8 \pi^2} A_0 Y_{\ell_i} \left( Y_\nu^\dagger 
Y_\nu \right)_{ij} {\rm Ln} \left( {M_{GUT} \over M} \right).\nonumber\\
\label{leading}
\end{eqnarray}
In this case of approximately degenerate heavy singlet neutrinos with a 
common mass $M \ll M_{GUT}$, as already mentioned, there is a 
single analogue of the Jarlskog invariant of the Standard 
Model~\cite{Jarlskog}:
\begin{equation}
J_{\tilde L} \; \equiv \; {\rm Im} \left[ \left( m_{\tilde L}^2 
\right)_{12} \left( m_{\tilde L}^2
\right)_{23} \left( m_{\tilde L}^2
\right)_{31} \right],
\label{J}
\end{equation}
which depends on the single phase that is observable in this 
approximation. There are other Jarlskog invariants defined analogously in 
terms of various combinations with the $A_\ell$, but these are all 
proportional~\cite{EHLR}.

There are additional contributions if the heavy singlet neutrinos are not 
degenerate, which contain the matrix factor
\begin{equation}
Y^\dagger L Y \; = \; X Y^d P_2 {\bar Z}^T L {\bar Z}^* P_2^* y^d 
X^\dagger,
\label{YLY}
\end{equation}
which introduces dependences on the phases in ${\bar Z} P_2$, though not 
$P_1$. In this way, the renormalization of the soft supersymmetry-breaking 
parameters becomes sensitive to a total of 3 CP-violating 
phases~\cite{EHRS}.

\section{Could the Inflaton be a Sneutrino?}

This `old' idea~\cite{MSYY1,MSYY2} has recently been
resurrected~\cite{ERY}. We recall that seesaw models~\cite{seesaw} of
neutrino masses involve three heavy singlet right-handed neutrinos
weighing around $10^{10}$ to $10^{15}$~GeV, which certainly includes the
preferred inflaton mass found above (\ref{phimass}). In addition, singlet
(s)neutrinos have no interactions with vector bosons, and have no cubic or
higher-order interactions in the minimal seesaw model. Hence the effective
potential for each sneutrino ${\tilde N}$ is simply $V = 1/2 |{\tilde
N}|^2$, as in the $\phi^2$ inflation model discussed earlier.

\begin{figure}
\centerline{\epsfxsize = 0.50\textwidth \epsffile{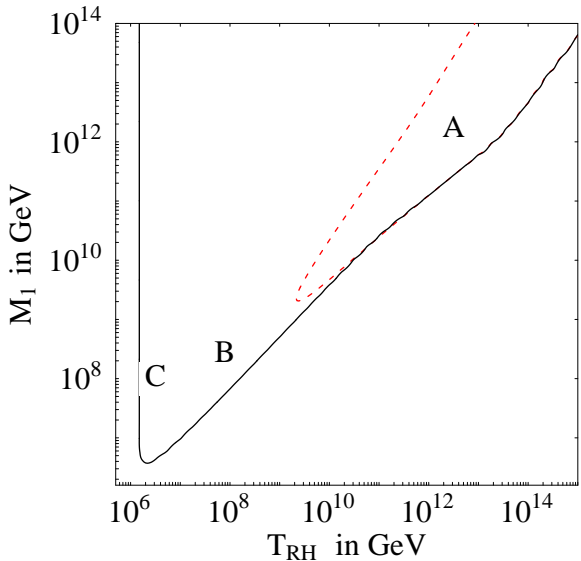}
}
\caption{
The solid curve bounds the region allowed for leptogenesis in the
$(T_{RH},\,M_{N_1})$ plane, assuming a baryon-to-entropy ratio $Y_B >
7.8\times
10^{-11}$ and the maximal CP asymmetry $\epsilon_1^{max}(M_{N_1}).$ 
Leptogenesis is entirely thermal in the
area bounded by the (red) dashed curve~\protect\cite{ERY,Strumia}.
}
\label{fig:reheat}
\end{figure}

Moreover, the Yukawa interaction $Y_\nu$ of the sneutrino is eminently
suitable for converting the inflaton energy density into particles via $N
\to H + \ell$ decays and their supersymmetric variants. 
Inflation ends when the Hubble 
expansion rate $H \sim m_{\tilde N}$, and the sneutrino inflaton field 
then oscillates around its minimum with an energy density that evolves 
like non-relativistic matter:
\begin{equation}
\rho_{\tilde N} \sim \rho_I \left( \frac{a_I}{a} \right)^3.
\label{nonrel}
\end{equation}
The oscillations continue until the inflaton decays, when the Hubble
expansion rate becomes comparable with the sneutrino decay rate: $H 
\sim \Gamma_{\tilde N}$. The sneutrino decay products then thermalize 
rapidly, reheating the Universe to a temperature $T_{RH}$ given by
\begin{equation}
T_{RH}^4 \propto \left( \frac{g^2_{\tilde N}}{8 \pi} m_{\tilde N} 
\right)^2 m_P^2.
\label{TRH}
\end{equation}
Since the
magnitudes of these Yukawa interactions are not completely determined,
there is flexibility in the reheating temperature after inflation, as we
see in Fig.~\ref{fig:reheat}~\cite{ERY}. Leptogenesis may be driven by 
a CP-violating asymmetry in the decays of the sneutrino inflaton. 
However, low reheating temperatures are 
favoured by considerations of the gravitino problem~\cite{gravitino}, as 
we discuss next, suggesting that leptogenesis is not thermal.

\section{THE GRAVITINO PROBLEM}

In models based on supergravity, the abundance of the gravitino ${\tilde
G}$ presents a problem~\cite{gravitino}, whether or not it is the lightest 
sparticle. Let us now consider these two different cases.

\subsection{The Gravitino is {\it not} the Lightest Sparticle}

In this case, the gravitino is unstable, and the lifetime for the 
simplest radiative decay into the lightest sparticle, assumed to the 
lightest neutralino $\chi$, is: 
\begin{equation}
\tau_{{\tilde G} \to \chi \gamma} \; \sim 3 \times 10^8 \times \left( 
\frac{100~{\rm GeV}}{m_{\tilde G}} \right)^3~{\rm s}.
\label{taugrav}
\end{equation}
There is a tight limit on the gravitino abundance before decay that is 
imposed by the consistency of the cosmological light-element abundances 
with the values calculated on the basis of the baryon-to-photon ratio 
inferred from CMB measurements, which might have been 
altered by 
photo-dissociation and other reactions. For a typical $\tau_{\tilde G} 
= 10^8$~s, one has~\cite{CEFO}:
\begin{equation}
Y_{\tilde G} \equiv \frac{n_{\tilde G}}{n_\gamma} \; < \; 5 \times 
10^{-14} \times \left( \frac{100~{\rm GeV}}{m_{\tilde G}} \right).
\label{Ygrav}
\end{equation}
This must be compared with the rate for thermal gravitino production 
following reheating after inflation~\cite{Buchgrav}:
\begin{equation}
Y_{\tilde G} \; \ga \; 10^{-11} \times \left( \frac{T_{RH}}{10^{10}~{\rm 
GeV}} \right),
\label{uppergrav}
\end{equation}
leading to the strong constraint~\cite{CEFO}
\begin{equation}
T_{RH} \; \la \; {\rm few} \times 10^7~{\rm GeV}.
\label{gravLSP}
\end{equation}
This would be strengthened by several orders of magnitude if hadronic 
gravitino decays become important~\cite{KMM}.

\subsection{The Gravitino {\it is} the Lightest Sparticle}

\begin{figure}
\begin{center}
\mbox{\epsfig{file=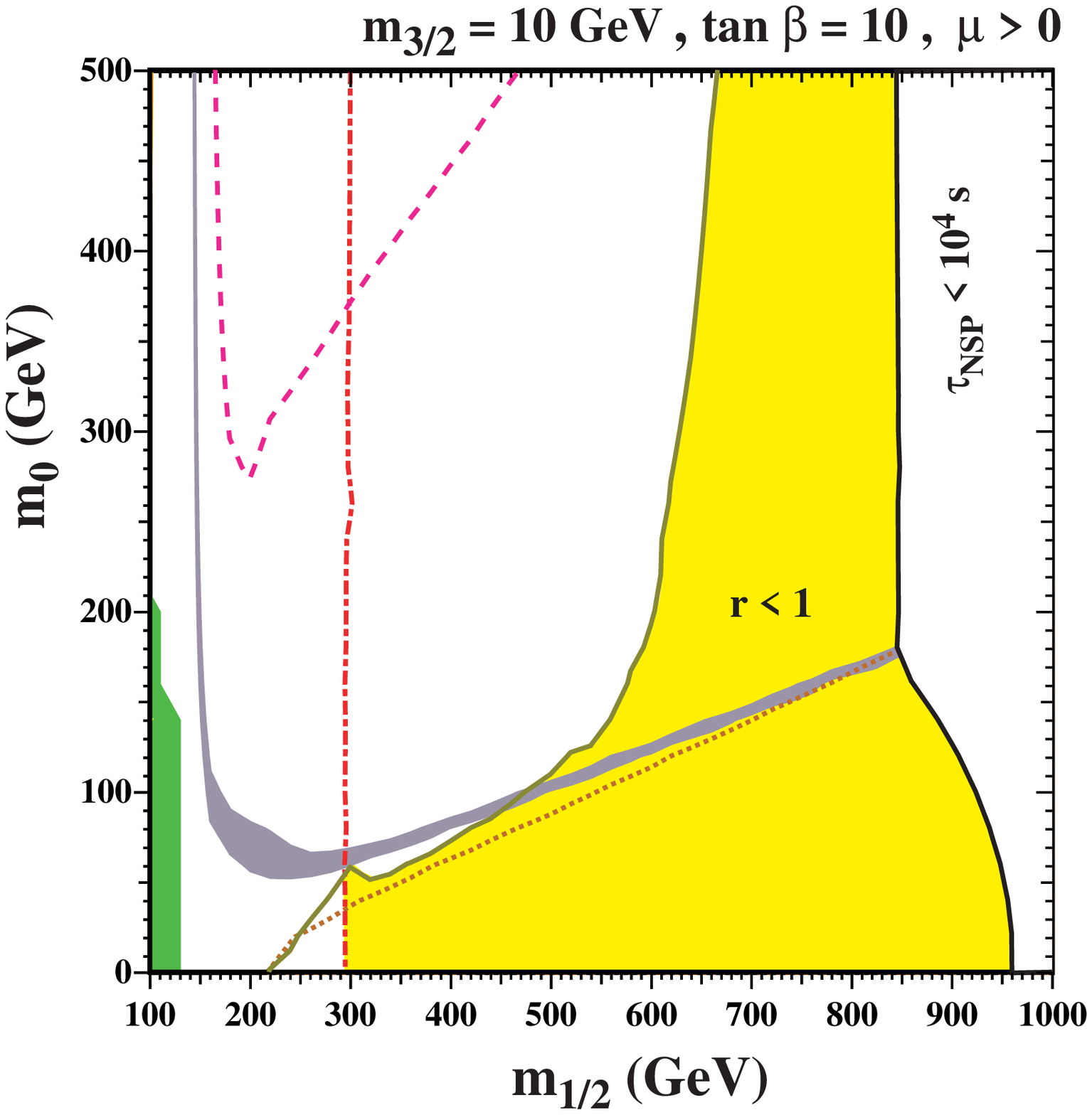,height=6cm}}
\mbox{\epsfig{file=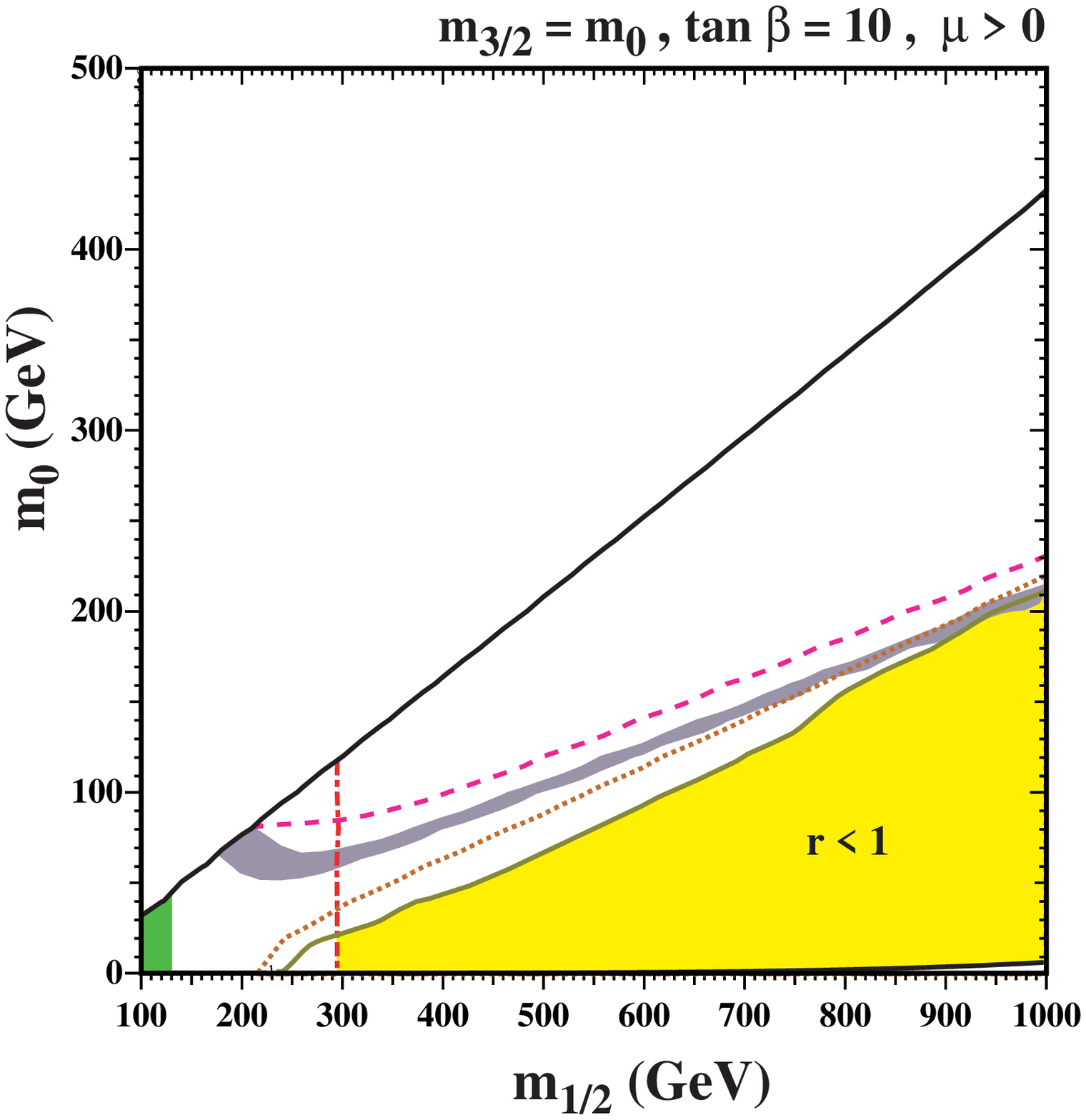,height=6cm}}
\end{center}
\caption{
The $(m_{1/2}, m_0)$ planes for $\tan \beta =10, \mu > 0$ and the choices
(a) $m_{\tilde G} = 10$~GeV, (b) $m_{\tilde G} = m_0$, assuming that the 
gravitino is the LSP~\protect\cite{EOSS}. In each panel, we show $m_h = 
114$~GeV calculated using {\tt FeynHiggs}~\cite{FeynHiggs}, as a near-vertical
(red)  dot-dashed line, the region excluded by $b \to s \gamma$ is darkly 
shaded(green), and the region where the NSP density before decay lies in 
the range $0.094 < \Omega^0_{NSP} h^2 < 0.129$ is medium shaded 
(grey-blue). The (purple) dashed line is the contour where gravitinos
produced in NSP decay have $\Omega_{3/2} h^2 = 0.129$, and the grey
(khaki) solid line ($r=1$) is the constraint on NSP decays provided by 
Big-Bang nucleosynthesis and CMB observations. The light (yellow) shaded 
region is allowed by all the constraints.  The contour where $m_\chi =
m_{\tilde \tau_1}$ is shown as a (red) diagonal dotted line. Panel (a) 
shows as a black solid line the contour beyond which $\tau_{NSP} \la
10^4$~s, a case not considered here. Panel (d) shows
black lines to whose left the gravitino is no longer the LSP.}
\label{fig:NSP10p}
\end{figure}

In this case, one must consider two important contributions to the relic 
gravitino abundance, from primordial production following inflationary 
reheating and from decays of the next-to-lightest sparticle (NSP), which 
might be the lightest neutralino $\chi$ or the lighter stau slepton ${\tilde 
\tau}_1$~\cite{EOSS}. In the latter mechanism, the NSP lifetime is 
typically
\begin{equation}
\tau_{NSP} \; \sim \; 10^4 \; {\rm to} \; 10^8~{\rm s},
\label{tauNSP}
\end{equation}
and one can recycle the previous light-element constraint on unstable 
relics to infer~\cite{EOSS}:
\begin{equation}
\Omega^0_{NSP} \; \la \; 10^{-2} \times \Omega_B h^2,
\label{ONSP}
\end{equation}
where $\Omega^0_{NSP}$ is the density the NSP would have had today, if the 
NSP had been stable. This is to be compared with standard calculations of 
relic sparticle abundances, which generally yield $\Omega_{LSP} \sim 5 
\Omega_B h^2$. Clearly, the requirement (\ref{ONSP}) is an important 
constraint on 
the supersymmetric model parameters. Indeed, it is generically much 
stronger than simply requiring $\Omega_{\tilde G} \le \Omega_{CDM}$, as 
illustrated in Fig.~\ref{fig:NSP10p} for the particular case of $\tan 
\beta = 10, \mu > 0$.

In this case, the constraint on the relic gravitino density from 
primordial thermal production is correspondingly weaker. For 
$\Omega_{\tilde G} \la 0.1$ as required by the astrophysical cold dark 
matter density, and $Y_{\tilde G} \ga 10^{-11} \times ( T_{RH} / 
10^{10}~{\rm GeV} )$, so that the reheating temperature is bounded 
by~\cite{EOSS}
\begin{equation}
T_{RH} \; \la 10^{10}~{\rm GeV} \times \left( \frac{100~{\rm 
GeV}}{m_{\tilde 
G}} \right).
\label{GLSP}
\end{equation}
Although this is considerably larger than the bound in the case when the 
gravitino is not the lightest sparticle, it is still much lower than the 
expected inflaton sneutrino mass.

\subsection{Implications for Leptogenesis}

Some can be inferred from Fig.~\ref{fig:reheat}, where the regions of
thermal and non-thermal leptogenesis are delineated~\cite{ERY}. If the
lightest sneutrino were lighter than about $10^{10}$~GeV, there would be a
small region where the gravitino LSP constraint (\ref{GLSP}) could be
satisfied.  However, the stronger gravitino NSP constraint is never
compatible with thermal leptogenesis. Moreover, if the lightest sneutrino
is responsible for inflation, with a mass $\sim 2 \times 10^{13}$~GeV as
estimated earlier, leptogenesis would have to be non-thermal, in both the
gravitino LSP and NSP scenarios described above.

\subsection{Implications for LFV}

\begin{figure*}[t]
\centerline{\epsfxsize = 0.5\textwidth \epsffile{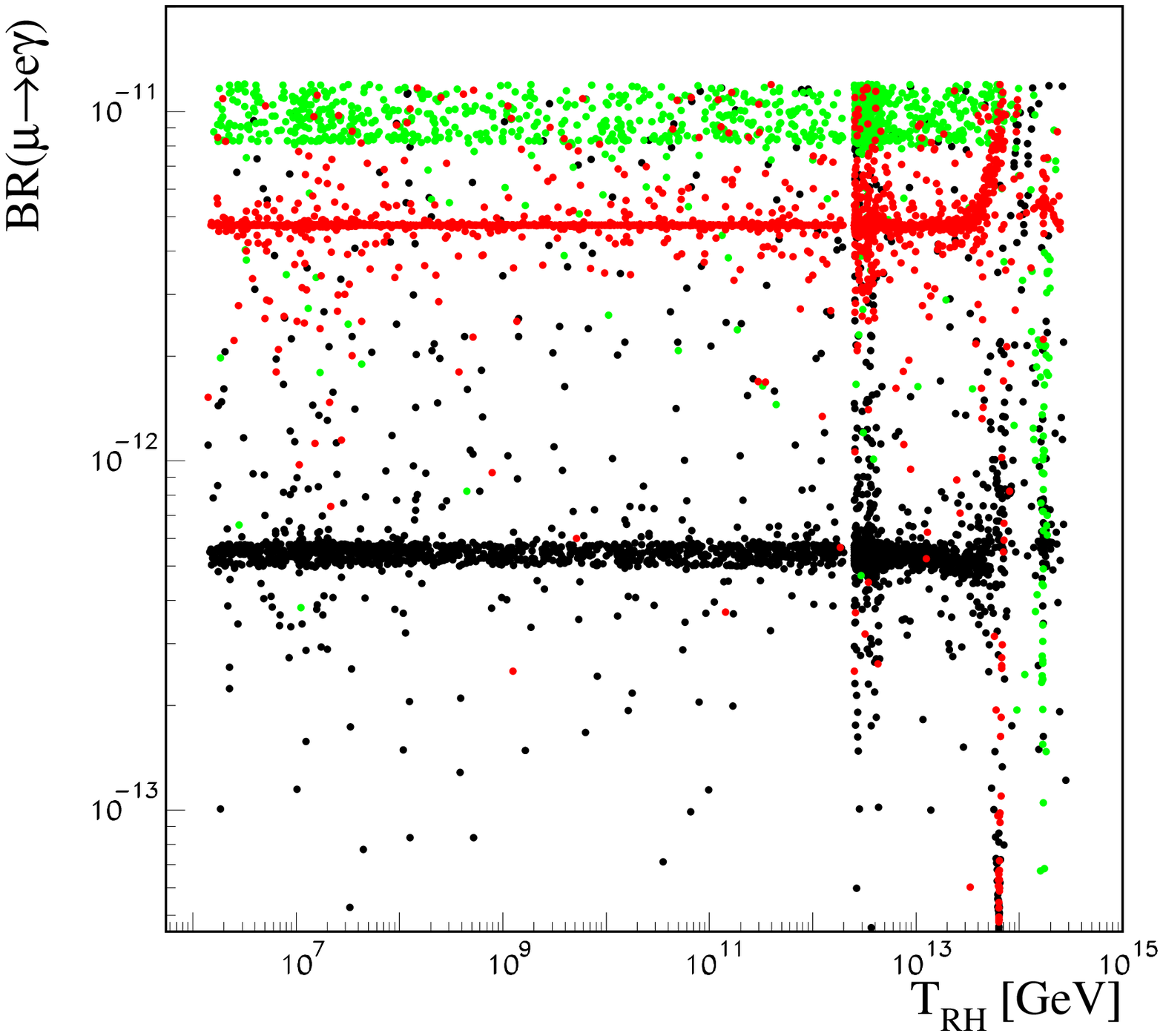}
\hfill \epsfxsize = 0.5\textwidth \epsffile{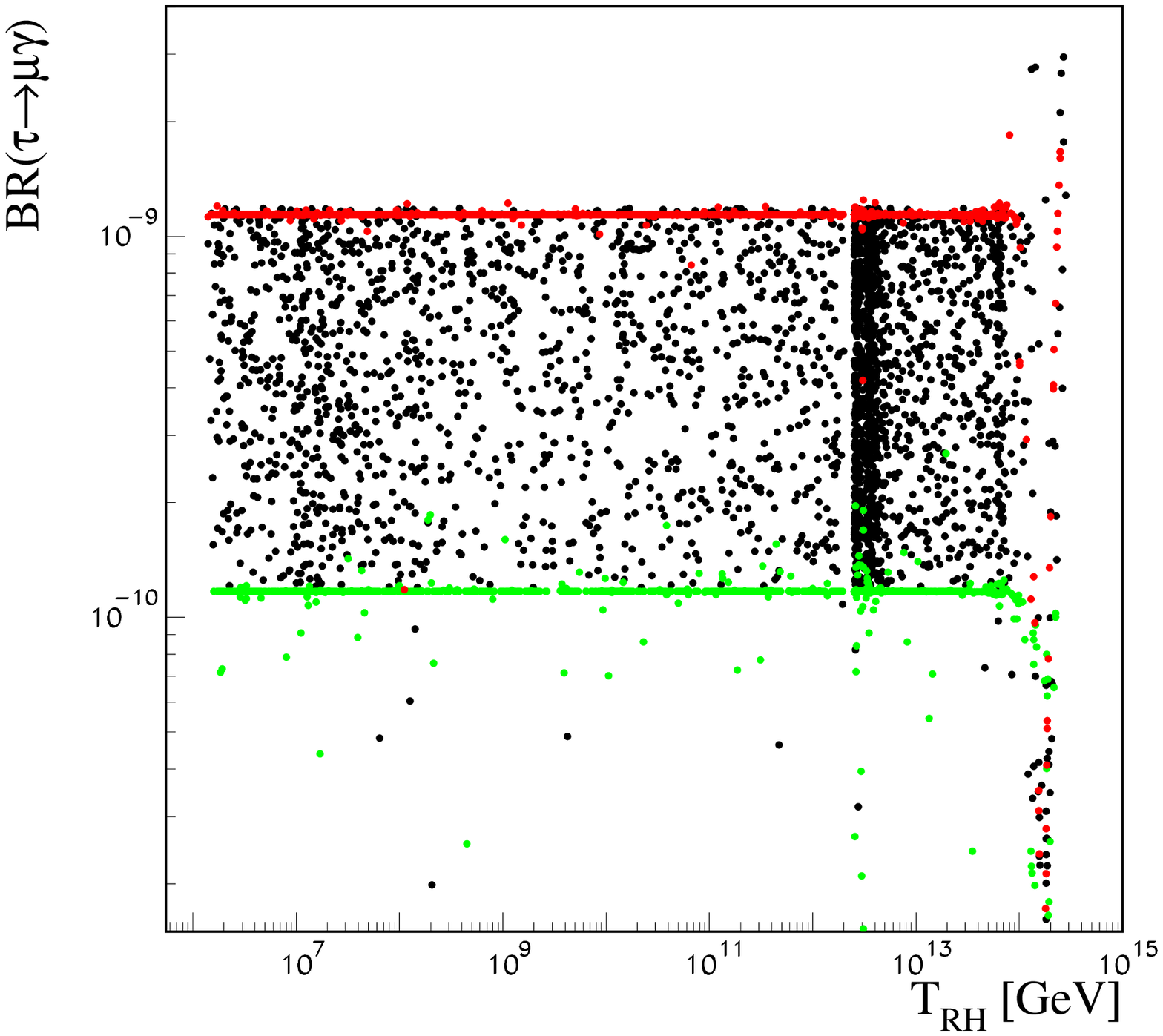}
}
\caption{
Calculations of BR$(\mu \to e \gamma)$ and BR$(\tau \to \mu \gamma)$   
in the left and right panels, respectively~\cite{ERY}. Black points 
correspond 
to $\sin \theta_{13} = 0.0$,  $M_2 = 10^{14}$~GeV,
and $5 \times 10^{14}$~GeV $< M_3 < 5 \times 10^{15}$~GeV.
Red points correspond to $\sin \theta_{13} = 0.0$,  $M_2 = 5 \times 
10^{14}$~GeV and  $M_3 = 5 \times 10^{15}$~GeV, while green points 
correspond to $\sin \theta_{13} = 0.1$,
$M_2 = 10^{14}$~GeV, and $M_3  = 5 \times 10^{14}$~GeV.
\vspace*{0.5cm}}
\label{fig3}
\end{figure*}

The hypothesis of sneutrino inflation constrains significantly the extra
parameters in the seesaw model. One of the heavy singlet (s)neutrino
masses is fixed at $\sim 2 \times 10^{13}$~GeV and, in the simplest case,
the other two must be heavier. Moreover, in order for the reheating
temperature after inflation to be acceptably low, the couplings of the
inflaton sneutrino must be quite small. Fig.~\ref{fig3} shows the
implications for the LFV decays $\mu \to e \gamma$ and $\tau \to \mu
\gamma$~\cite{ERY}. In both cases, we see that the LFV decays are
essentially independent of $T_{RH}$ once it is less than about
$10^{12}$~GeV, so the predictions would be similar in the gravitino LSP
and NSP scenarios decsribed above.  Comparing the top and bottom bands in
the first panel, we see that the branching ratio of $\mu \to e \gamma$ has
considerable sensitivity to $\sin \theta_{13}$ and, comparing the bottom
two bands, also to the mass $M_3$ of the heaviest singlet neutrino. In the
second panel, we see that the branching ratio for $\tau \to \mu \gamma$
depends differently on $\sin \theta_{13}$ and is much more sensitive to
$M_3$.

\section{LFV IN DECOUPLING MODELS}

Sneutrino inflation is one hypothesis motivating the (near-)decoupling of
one heavy singlet sneutrino, which is also motivated by some flavour
models of neutrino masses within the seesaw model~\cite{Chank}. In this
Section, we explore in more detail the implications of such decoupling for
LFV processes~\cite{CEPRT}.

We may write the neutrino Yukawa coupling matrix in the general form
\begin{equation}
Y_\nu \; \propto M_N^{1/2} \Omega U_\nu,
\label{decomp}
\end{equation}
where $U_\nu$ is the MNS neutrino mixing matrix:
\begin{eqnarray}
{U}_\nu&=&\left(\matrix{
c_{12} & s_{12} & s_{13}e^{-i\delta}\cr
-\frac{s_{12}}{\sqrt{2}}+\dots &
\frac{c_{12}}{\sqrt{2}}+\dots & {1\over\sqrt2} \cr
{s_{12}\over\sqrt{2}}+\dots & -{c_{12}\over\sqrt{2}}+\dots &
{1\over\sqrt2}} \right) \nonumber \\
&\cdot&\mathrm{diag}(e^{i\phi_1},e^{i\phi_2},1),
\end{eqnarray}
where neutrino data suggest that $\sin^2 \theta_{13} \sim 0.315$, the 
Majorana phases $\phi_{1,2}$ are unknown, and we consider two extreme 
hypotheses for $\theta_{13}$ and the CP-violating phase 
$\delta$~\cite{CEPRT}:
\begin{equation}
{\rm a):} \; \sin \theta_{13} = 0, \; {\rm b):} \; \sin \theta_{13} = 0.1, 
\delta = \frac{\pi}{2}.
\label{theta13}
\end{equation}
Furthermore, we assume
\begin{equation}
m_{\nu_1} \ll m_{\nu_2} < m_{\nu_3}, \; M_1 \le M_2 < M_3.
\label{ineq}
\end{equation}
Finally, we assume that one of the flavours (almost) decouples from the 
other two in the unknown matrix $\Omega$. The (almost) decoupled flavour 
may be any one of the three:
\begin{eqnarray}
\label{omegalep}
\textrm{decoupling of $N_1$} & & \Omega_\nu =
\left(\begin{array}{ccc} 1 & 0 & 0 \\ 0 & z & p \\ 0 & \mp p & \pm z
\end{array}\right) \\
\label{omegatwo}
\textrm{decoupling of $N_2$} & & \Omega_\nu =
\left( \begin{array}{ccc} 0 & z & p \\ 1 & 0 & 0 \\ 0 & \mp p & \pm z
\end{array}\right) \\
\label{omegadec}
\textrm{decoupling of $N_3$} & & \Omega_\nu =
\left( \begin{array}{ccc} 0 & z & p \\ 0 & \pm p & \mp z \\ 1 & 0 & 0   
\end{array}\right)
\end{eqnarray}
where $z^2+p^2=1$. The first of these options was that explored 
in~\cite{ERY}, and will be studied further here: the other options were 
discussed in~\cite{CEPRT}.

To a good approximation, the LFV branching ratios for $\ell_A \to \ell_B +
\gamma$ processes are proportional to the off-diagonal soft
supersymmetry-breaking quantities $| {\tilde m}^2_{AB} |^2$, where in a 
leading-logarithmic approximation
\begin{equation}
\tilde m^2_{AB} \propto \Sigma_C(Y^{CA}_\nu)^* \ln(m_X/M_C)(Y^{CB}_\nu)~.
\label{approx}
\end{equation}
This is a useful guide to understanding the results shown below, which are 
however based on calculations using the full one-loop 
renormalization-group equations.

\subsection{$\tau \to \mu \gamma$}

In the first decoupling pattern (\ref{omegalep}), we have~\cite{CEPRT} 
\begin{eqnarray}
&&{\tilde m}^2_{AB}\, \approx  \,
\left[U_\nu^{33}U_\nu^{23\ast}\left(|z|^2 + S |1-z^2|\right) \right.+\nonumber\\
&+&  R \,U_\nu^{33}U_\nu^{22\ast}
\left(S z\sqrt{1-z^2}^\ast - z^\ast\sqrt{1-z^2}\right) + \nonumber \\
&+& R\,U_\nu^{32}U_\nu^{23\ast}\left(S z^\ast\sqrt{1-z^2} - 
z\sqrt{1-z^2}^\ast\right) + \nonumber \\
 & + &\left.R^2\,U_\nu^{32}U_\nu^{22\ast}\left( S |z|^2 + |1-z^2| \right)
\right],\nonumber \\
\label{q32}
\end{eqnarray}
where $R=\sqrt{m_{\nu_2}/m_{\nu_3}}\sim 0.41$ and
$S$ gives the subleading
contribution of the product $(Y_{\nu_{2A}})^* Y_{\nu_{2B}}$.
The branching ratio does not depend strongly on the masses $M_1$
and $M_2$ and the Majorana phase $\phi_1$, as long as $M_1<M_2\ll M_3$,
i.e., for $S\ll1$. For illustration, we take
$M_3=5\times10^{14}$~GeV, $M_2=3\times 10^{13}$~GeV ($S\approx 0.1$) and
$M_1=2\times 10^{13}$~GeV, consistent with inflation being driven by the
lightest singlet sneutrino.

\begin{figure}
\includegraphics*[height=8cm]{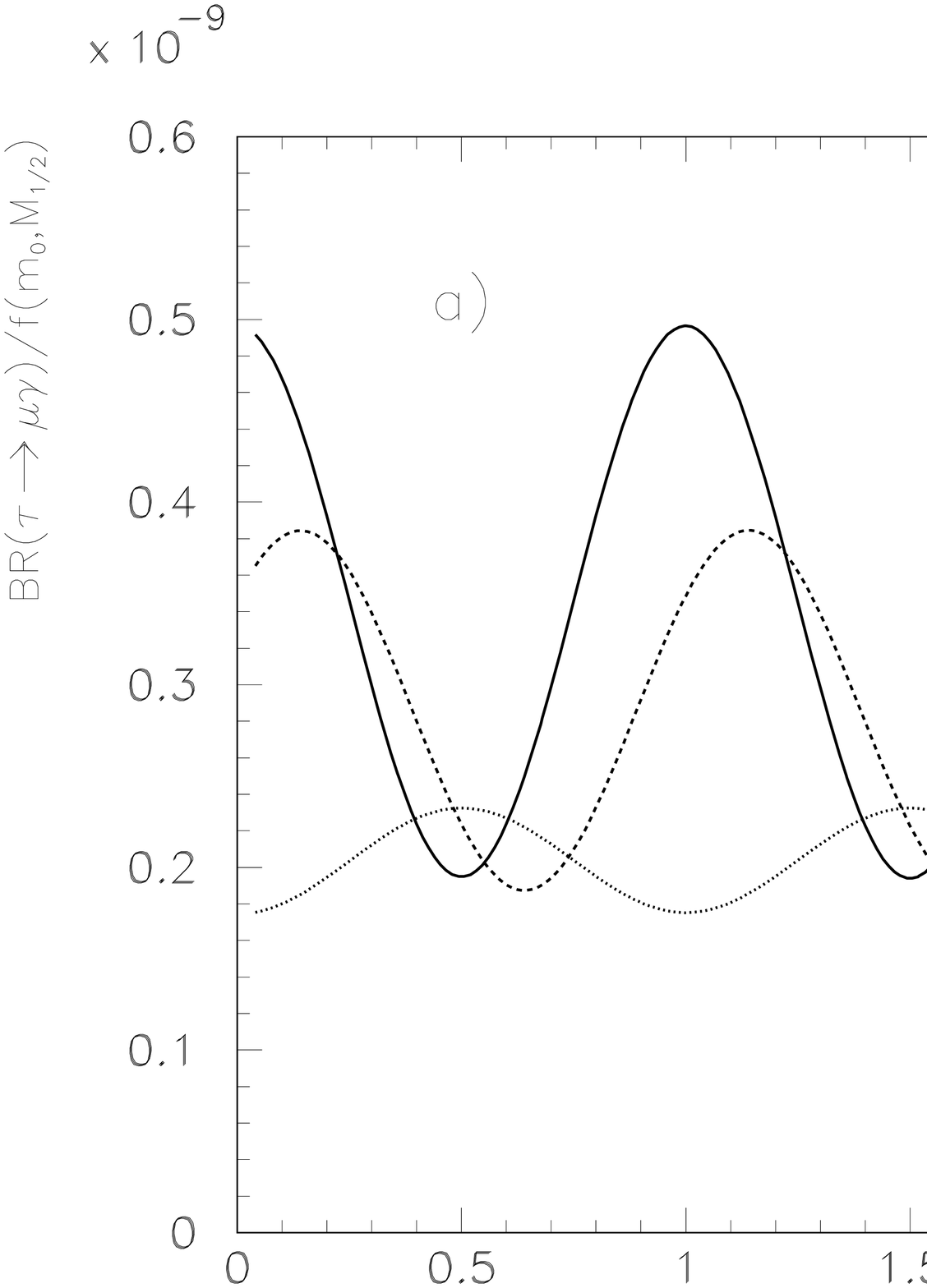}
\hspace{1cm}
\includegraphics*[height=8cm]{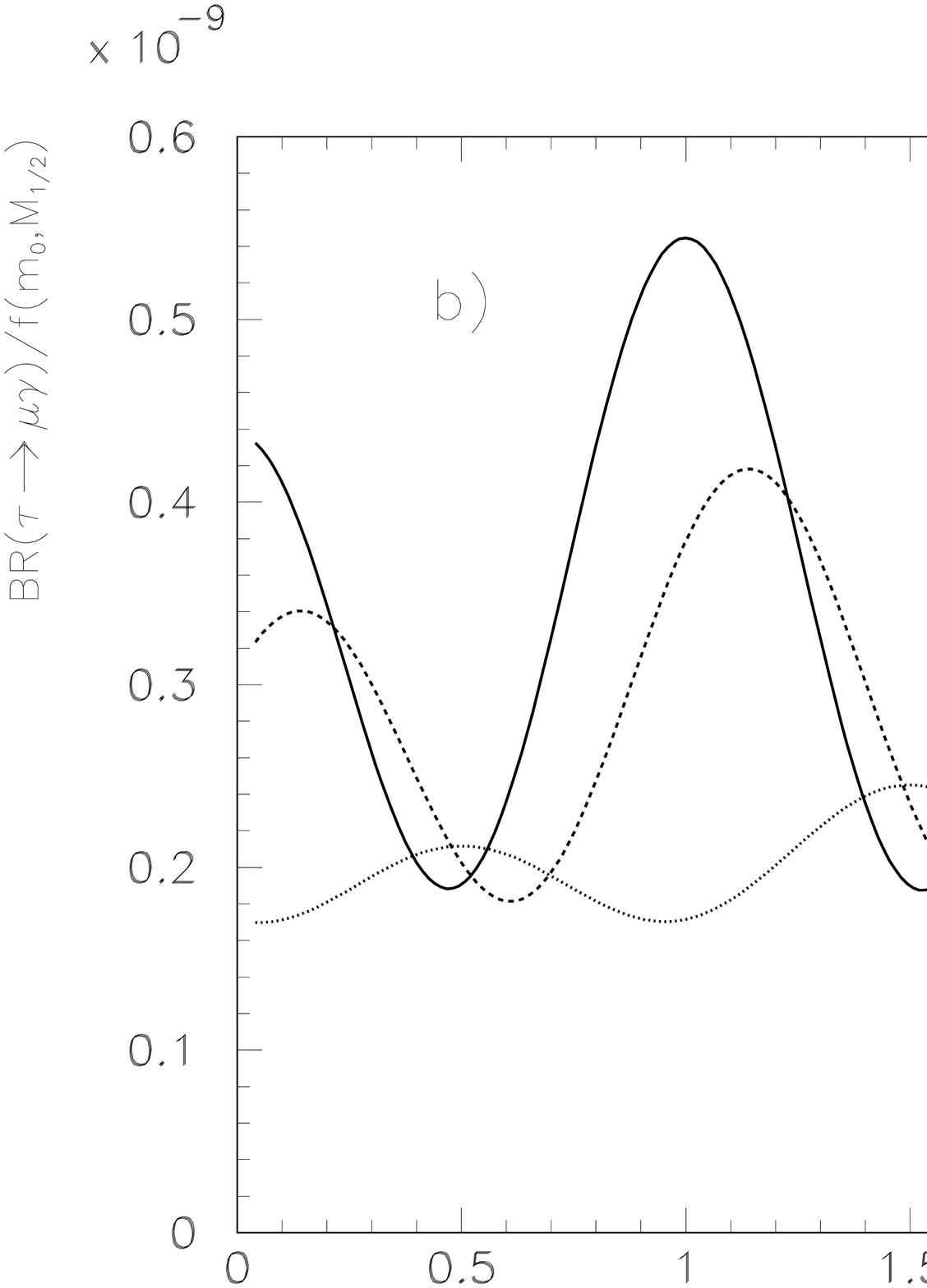}
\caption{ The branching ratio for $\tau\to\mu\gamma$ 
divided by a kinematic factor $f(m_0,M_{1/2})$, 
$BR(\tau\to\mu\gamma)/f(m_0,M_{1/2})$, as a function of
$\phi_2$ for the choices of $\theta_{13}, \delta$ in (\ref{theta13}),
$|z|=1/\sqrt2$, $\tan\beta=10$ and $A_0=0$~\protect\cite{CEPRT}. Dotted, 
dashed and solid lines correspond to $\arg z=0,\pi/4,\pi/2$, respectively. 
\label{fitmg1}}   
\end{figure}

\begin{figure}
\includegraphics*[height=8cm]{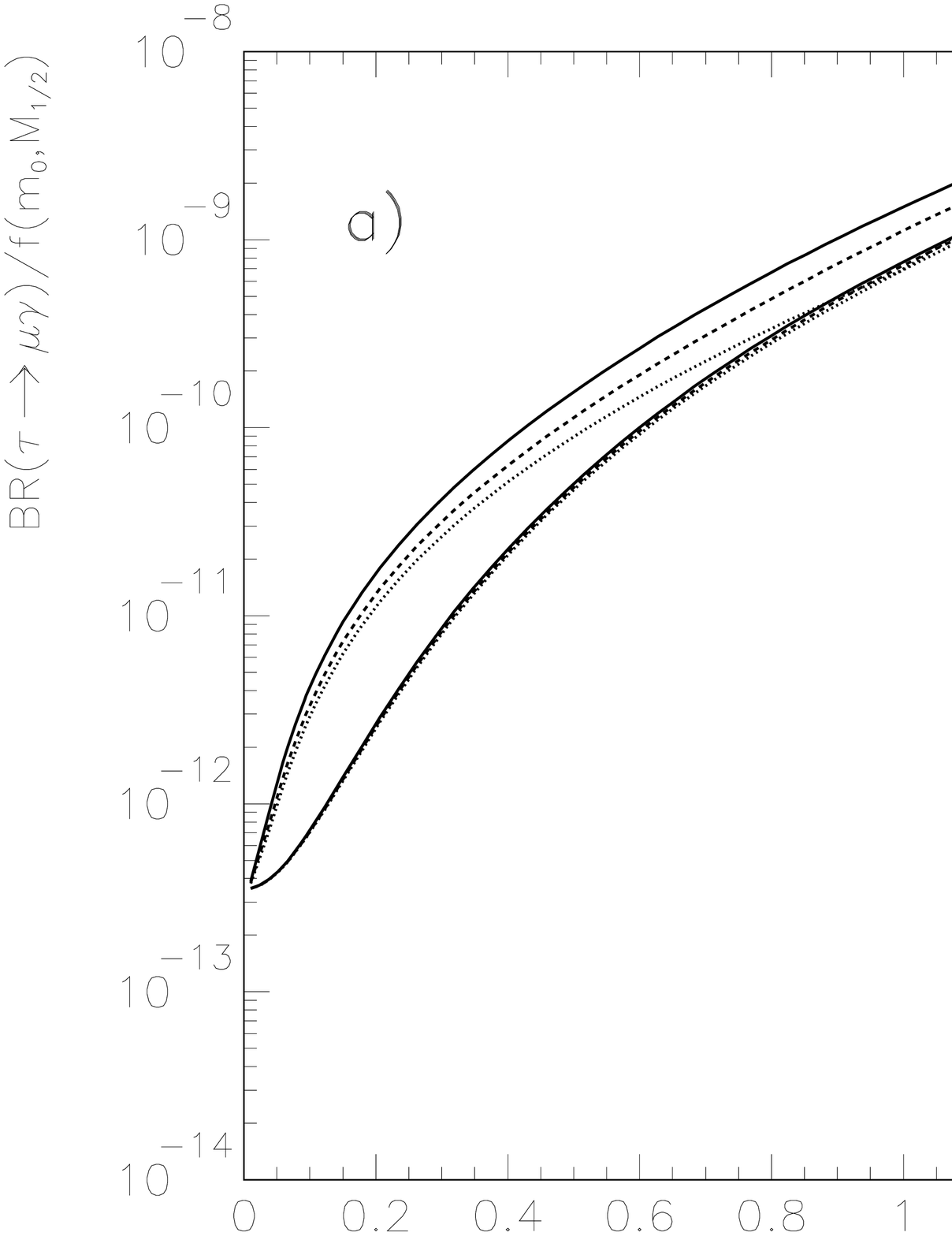}  
\hspace{1cm}
\includegraphics*[height=8cm]{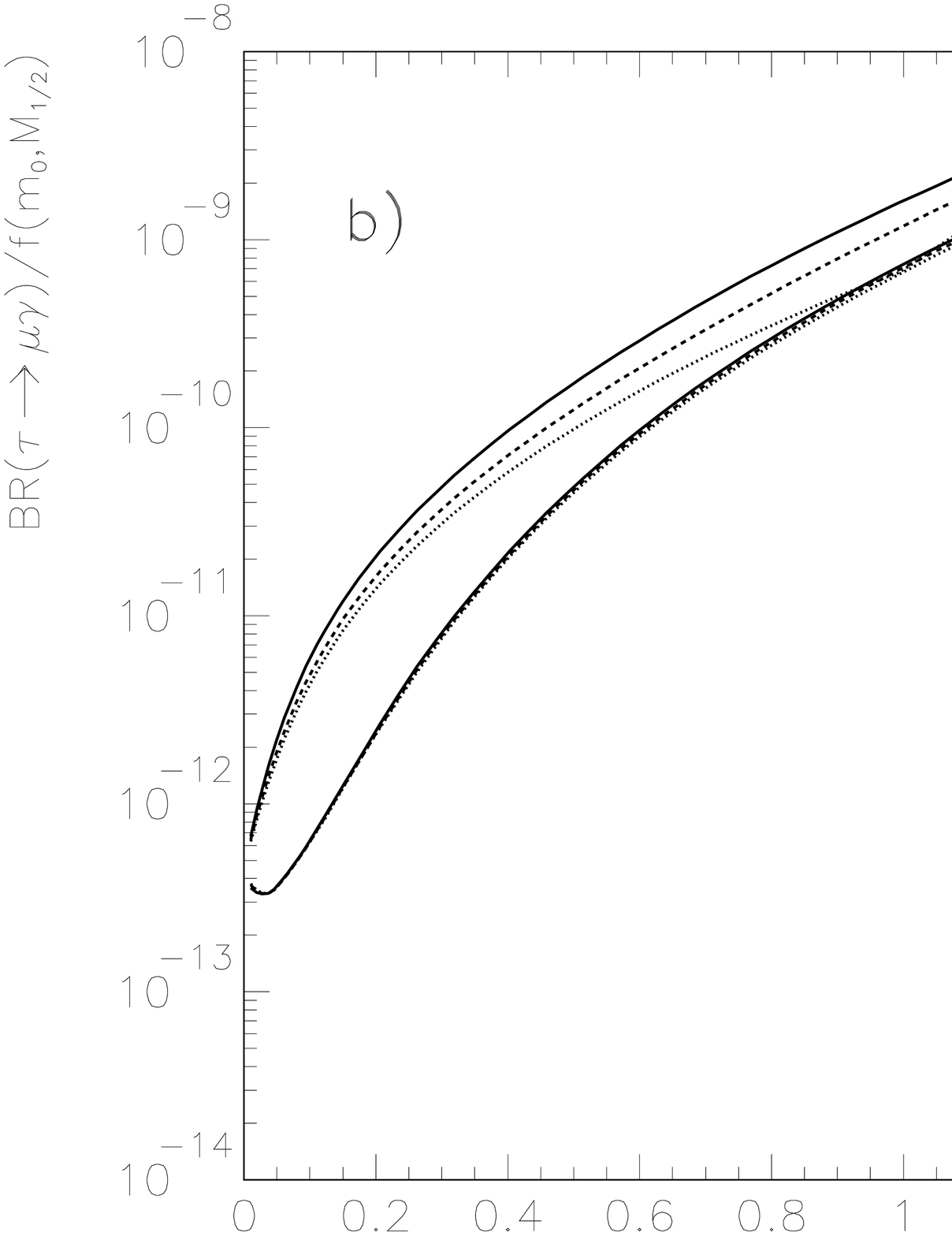}  
\caption{ Extremal values of $BR(\tau\to\mu\gamma)/f(m_0,M_{1/2})$ as  
a function of $|z|$ for the choices of $\theta_{13}, \delta$ in 
(\ref{theta13}), $\tan\beta =10$ and $A_0=0$~\protect\cite{CEPRT}. Dotted, 
dashed and solid lines correspond to $\arg z=0,\pi/4,\pi/2$, respectively. 
\label{fitmg2}}
\end{figure}

We see in Fig.~\ref{fitmg1} that the branching ratio for
$\tau\to\mu\gamma$ does not vary greatly with $\phi_2$ and the phase of
$z$, for representative choices of the other parameters and the two
options (\ref{theta13}) for $\theta_{13}, \delta$. On the other hand, we
see in Fig.~\ref{fitmg2} that the branching ratio for $\tau\to\mu\gamma$
does vary quite significantly with $|z|$~\cite{CEPRT}. It would seem from
these plots that $\tau\to\mu\gamma$ might have a branching ratio above
$10^{-9}$, but before reaching this conclusion we must examine the model
predictions for $\mu \to e \gamma$.

\subsection{$\mu \to e \gamma$}

\begin{figure}
\includegraphics*[height=8cm]{fi2tmga.ps}
\hspace{1cm}
\includegraphics*[height=8cm]{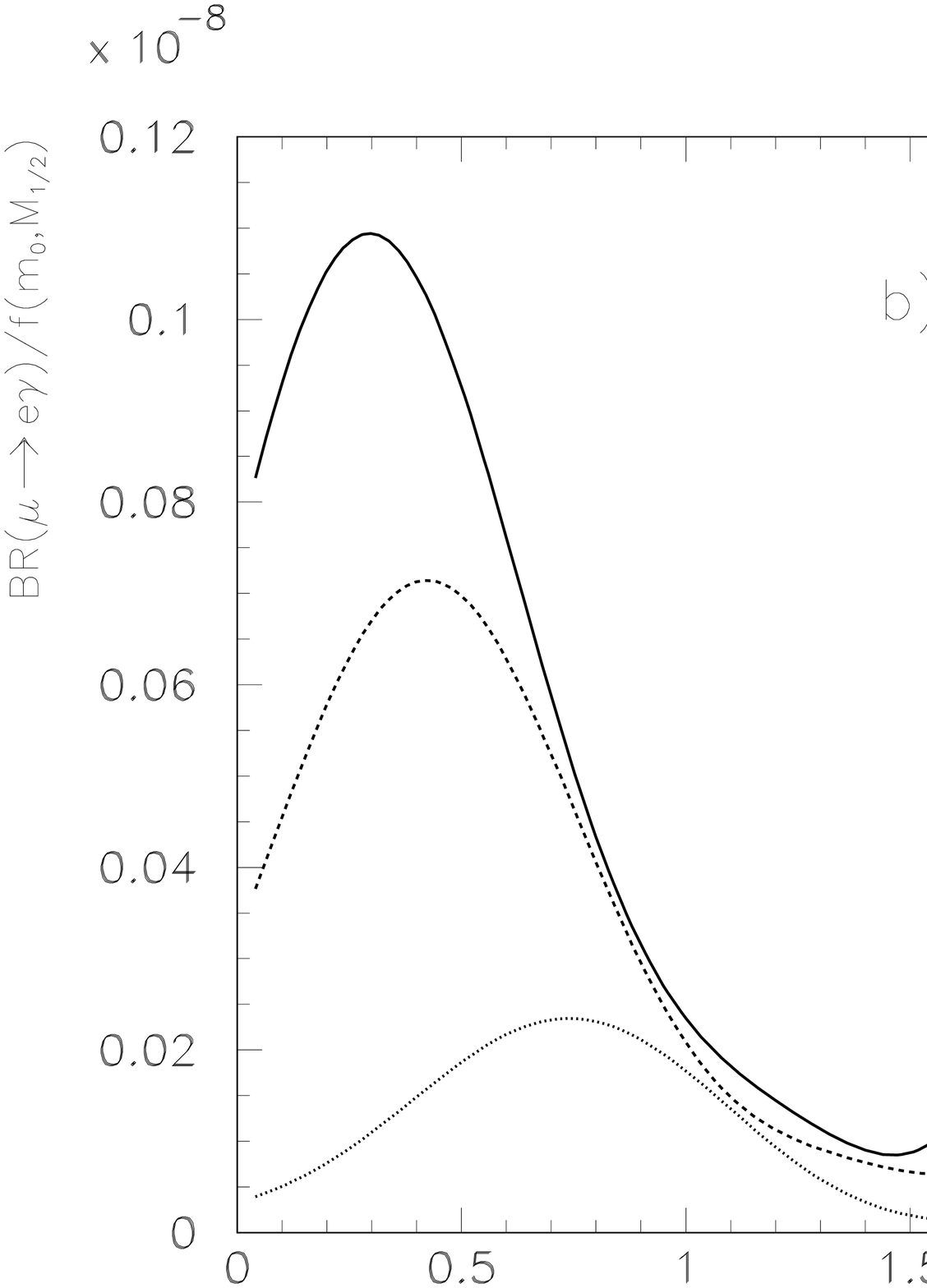}
\caption{ $BR(\mu\to e\gamma)/f(m_0,M_{1/2})$ as a function of $\phi_2$
for the choices of $\theta_{13}, \delta$ in (\ref{theta13}),
$|z|=1/\sqrt2$, $\tan\beta =10$ and $A_0=0$~\protect\cite{CEPRT}. Dotted, 
dashed and solid lines correspond to $\arg z=0$, $\pi/4$ and $\pi/2$, 
respectively. 
\label{fimeg1}}
\end{figure}

\begin{figure}
\includegraphics*[height=8cm]{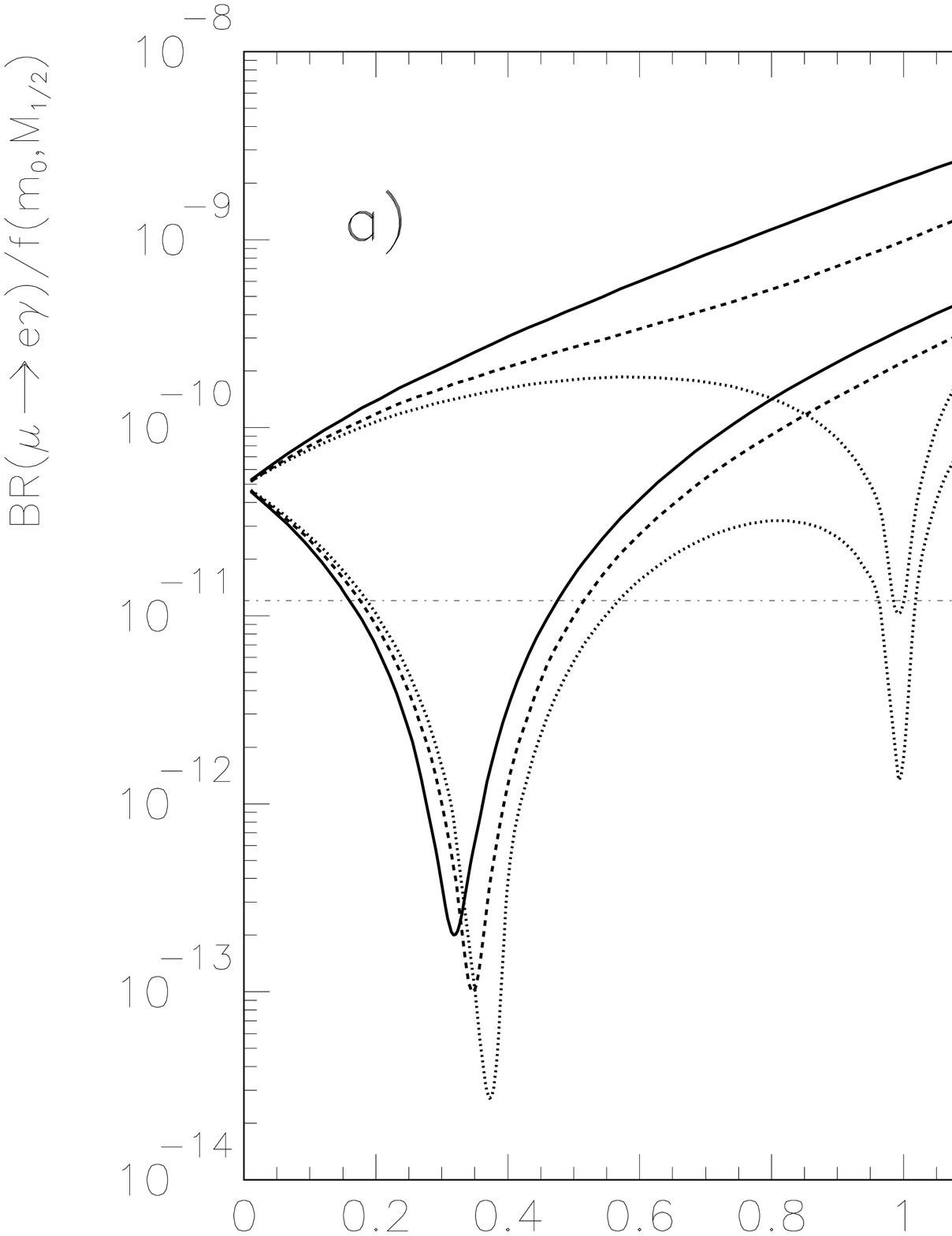}
\hspace{1cm}
\includegraphics*[height=8cm]{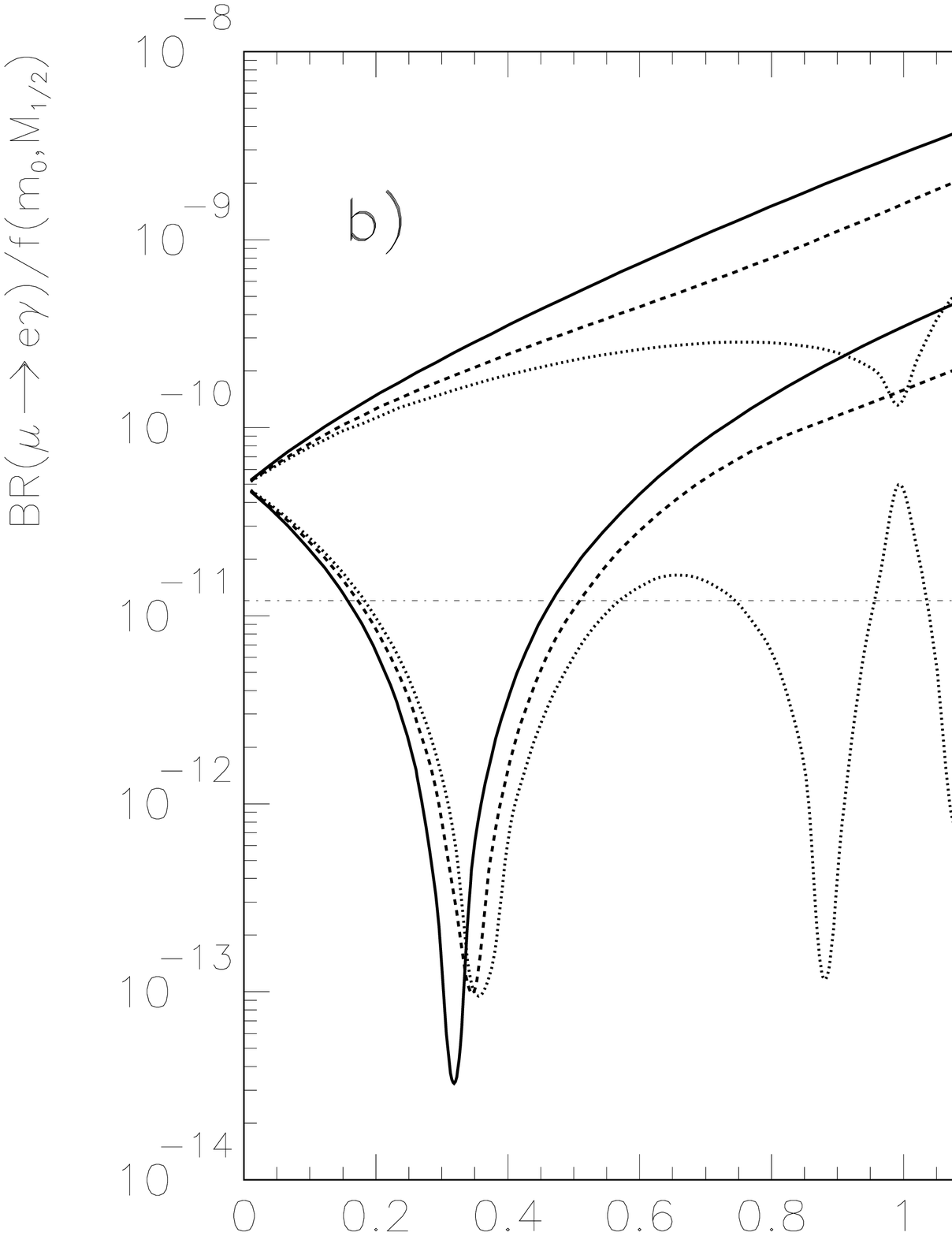}
\caption{ Extremal values of $BR(\mu\to e\gamma)/f(m_0,M_{1/2})$ as a
function of $|z|$ for the choices of $\theta_{13}, \delta$ in
(\ref{theta13}), $\tan\beta =10$ and $A_0=0$~\protect\cite{CEPRT}. Dotted, 
dashed and solid lines correspond to $\arg z=0$, $\pi/4$ and $\pi/2$, 
respectively. 
\label{fimeg2}}
\end{figure}

The quantity ${\tilde m}^2_{L_{21}}$ relevant for $\mu\to e\gamma$ decay
can be approximated by:
\begin{eqnarray}
&& {\tilde m}^2_{L_{21}} \propto  \nonumber\\
&&\left[R U_\nu^{A3}U_\nu^{12*}\left(Sz\sqrt{1-z^2}^\ast- 
z^\ast\sqrt{1-z^2}\right) \right.\nonumber\\
&+& R^2 U_\nu^{A2} U_\nu^{12\ast}\left(S|z|^2+|1-z^2|\right) + \nonumber \\
&+&\left.  U_\nu^{A3} U_\nu^{13\ast}\left(|z|^2+S|1-z^2|\right)\right],
\label{q21}
\end{eqnarray}
where we use again the decoupling texture (\ref{omegalep})~\cite{CEPRT}.
As seen in Fig.~\ref{fimeg1}, $\mu \to e \gamma$ exhibits a stronger 
dependence on $\phi_2$ than does $\tau \to \mu \gamma$ and, as seen in 
Fig.~\ref{fimeg2}, it exhibits cancellations for some specific values of 
$|z|$ and the other parameters. This is just as well, because the 
prediction for the branching ratio of $\mu \to e \gamma$ rises above the 
present experimental limit of $1.2 \times 10^{-11}$ for generic values of 
the parameters. However, even in the narrow regions where there is a
cancellation, the branching ratio of $\mu \to e \gamma$ generally exceeds 
$10^{-13}$, within the sensitivity of the ongoing experiment at PSI. 
Unfortunately, referring back to Fig.~\ref{fitmg2}, we see that the 
regions where $\mu \to e \gamma$ is acceptably rare do not have large 
branching ratios for $\tau \to \mu \gamma$. This is a pity, as the ratio 
of the two decays would provide valuable information about a complex 
parameter of the seesaw model that is inaccessible in neutrino oscillation 
experiments~\cite{CEPRT}.

\subsection{$\tau \to e \gamma$}

\begin{figure}
\includegraphics*[height=8cm]{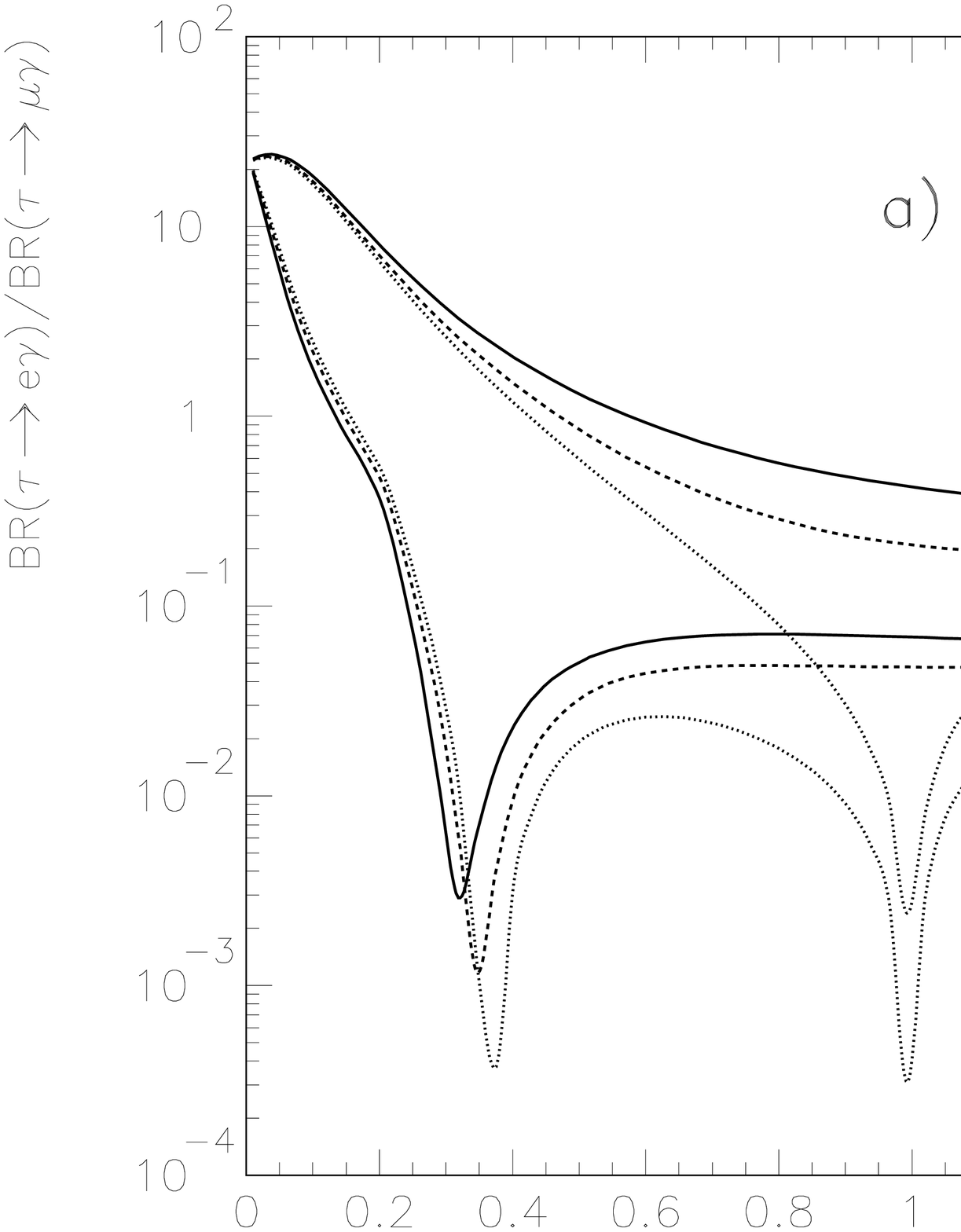}
\hspace{1cm}
\includegraphics*[height=8cm]{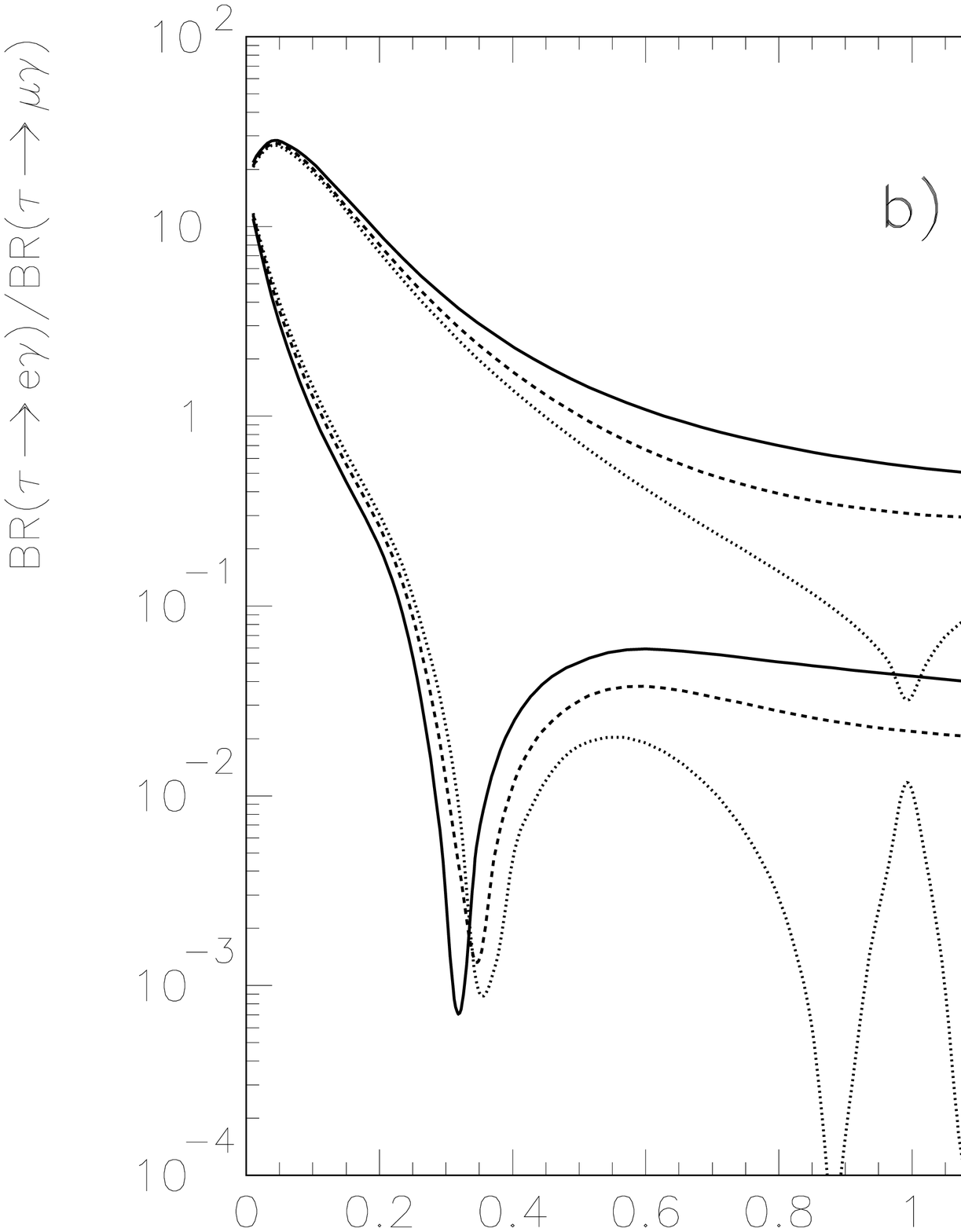}
\caption{The extremal values of $BR(\tau\to e\gamma)/BR(\tau\to
\mu\gamma)$
as a function of $|z|$ for the choices of $\theta_{13}, \delta$ in 
(\ref{theta13}),
$\tan\beta =10$ and $A_0=0$~\protect\cite{CEPRT}. Dotted, dashed and solid 
lines correspond to $\arg z=0$, $\pi/4$ and $\pi/2$, respectively. 
\label{fiteg2}}
\end{figure}

We consider finally the decay $\tau \to e \gamma$. Within the decoupling
texture (\ref{omegalep}), this has a dependence on $\phi_2$ that is quite
strong but opposite to that of $\tau \to \mu \gamma$. The branching ratio
is generally smaller than that for $\tau \to \mu \gamma$, as seen in
Fig.~\ref{fiteg2}, particularly in the cancellation regions~\cite{CEPRT}.

\section{CONCLUSIONS}

The contemporary techno-folk singer Moby tells us that `We are all made of
stars'. Perhaps this is a reference to astrophysical nucleosynthesis? In
contrast, my main message in this talk is that `We are all made of
neutrinos', not just in the sense of leptogenesis, but also in the sense
that the overall size, age, flatness and entropy are all due to neutrinos,
via cosmological inflation. Moreover, this hypothesis enables some
testable predictions to made for the radiative charged-lepton decays $\mu
\to e \gamma$, $\tau \to \mu \gamma$ and $\tau \to e \gamma$.

\end{document}